\documentclass[%
reprint,
superscriptaddress,https://www.overleaf.com/project/5dc3febdc6430c00018feb6d
linenumbers,
nofootinbib,
longbibliography,
amsmath,amssymb,
aps,
pra,
]{revtex4-2}

\usepackage{graphicx}
\usepackage{dcolumn}
\usepackage{bm}

\usepackage{color,colortbl}
\usepackage{units}
\usepackage{tikz}
\usetikzlibrary{arrows,shapes}

\usepackage{algorithm}
\makeatletter
\renewcommand{\ALG@name}{Table 1}
\makeatother

\usepackage[noend]{algpseudocode}
\usepackage{romannum}

\usepackage{diagbox} 

\newcolumntype{C}[1]{>{\centering\let\newline\\\arraybackslash\hspace{0pt}}m{#1}}

\usepackage[caption=false]{subfig}
\usepackage{float}

\definecolor{mygreen}{rgb}{0.1, 0.6, 0.0}

\definecolor{pap}{rgb}{0.54, 0.17, 0.89}

\usepackage{blindtext}
\usepackage[inline]{enumitem}

\usepackage{hyperref}

\DeclareMathAlphabet\mathbfcal{OMS}{cmsy}{b}{n}

\usepackage{ulem}

\usepackage{soul}
\usepackage{xcolor,cancel}

\begin{document}


\title{High-Q cavity coupled to a high permittivity dielectric resonator for sensing applications}

	\author{Shahnam Gorgi Zadeh}
\email{shahnam.zadeh@uni-rostock.de}
	\affiliation{European Organization for Nuclear Research (CERN), Meyrin 1217, Switzerland}
 
	\author{Alberto Ghirri}
 \email[Corresponding author. ]{alberto.ghirri@nano.cnr.it}
	\affiliation{Istituto Nanoscienze - CNR, Centro S3, via G. Campi 213/A, 41125, Modena, Italy}

     \author{Sergio Pagano}
	\affiliation{Physics Department, University of Salerno and INFN GC Salerno, via Giovanni Paolo II 132, Fisciano (SA), Italy}

    \author{Simone Tocci}
	\affiliation{INFN, National Institute for Nuclear Physics, I-00044, Frascati, Italy}

    \author{Claudio Gatti}
	\affiliation{INFN, National Institute for Nuclear Physics, I-00044, Frascati, Italy}

 	\author{Antonio Cassinese}
	\affiliation{Physics Department and INFN -Napoli,  Università Napoli Federico II, P.le Tecchio 80, 80125 Naples, Italy}

\begin{abstract}
The use of high-quality factor resonators is of undoubted interest for high-precision measurements and for applications in quantum technologies. Novel types of microwave sensors can be realized by coupling a first resonator acting as a stable frequency reference with a second resonator that is sensitive to a particular physical quantity. Here we report on a coupled cavity configuration in which a high $Q$ factor elliptical TESLA-shaped superconducting cavity is coupled with a high permittivity ($\varepsilon_\mathrm{r}$) SrTiO$_3$ puck, whose resonant frequency varies as a function of temperature due to the temperature dependence of the permittivity that reaches values higher than 30000 below 1~K. Extensive electromagnetic simulations are used to test different coupling configurations, showing great versatility in tuning the coupling in the weak or strong regime, depending on the puck's position within the cavity. Moreover, for the coupled system, they allow investigation of the dependence of the zero transmission frequency (notch frequency) on changes in permittivity $\varepsilon_\mathrm{r}$, obtaining a maximum value of 2.8 MHz per unit change of $\varepsilon_\mathrm{r}$, for $\varepsilon_\mathrm{r} \approx 230$. Finally, we discuss the use of the coupled system as a sensor that operates in different temperature ranges.
		
\end{abstract}

\maketitle

	\section{Introduction}
	\label{sec:introduction}

Resonant cavities, with their ability to sustain high electromagnetic fields, have emerged as powerful tools for ultrasensitive measurements. These structures, also known as resonators, operate by confining electromagnetic waves within a bounded region, allowing resonance at specific frequencies. The high quality factors ($Q$ factors) associated with resonant cavities provide high sensitivity and therefore potential high measurement precision and accuracy for measurements of different physical quantities, making them indispensable in various scientific and engineering applications. 

    \begin{figure*}[t]
		\centering
        \includegraphics[width=\linewidth]{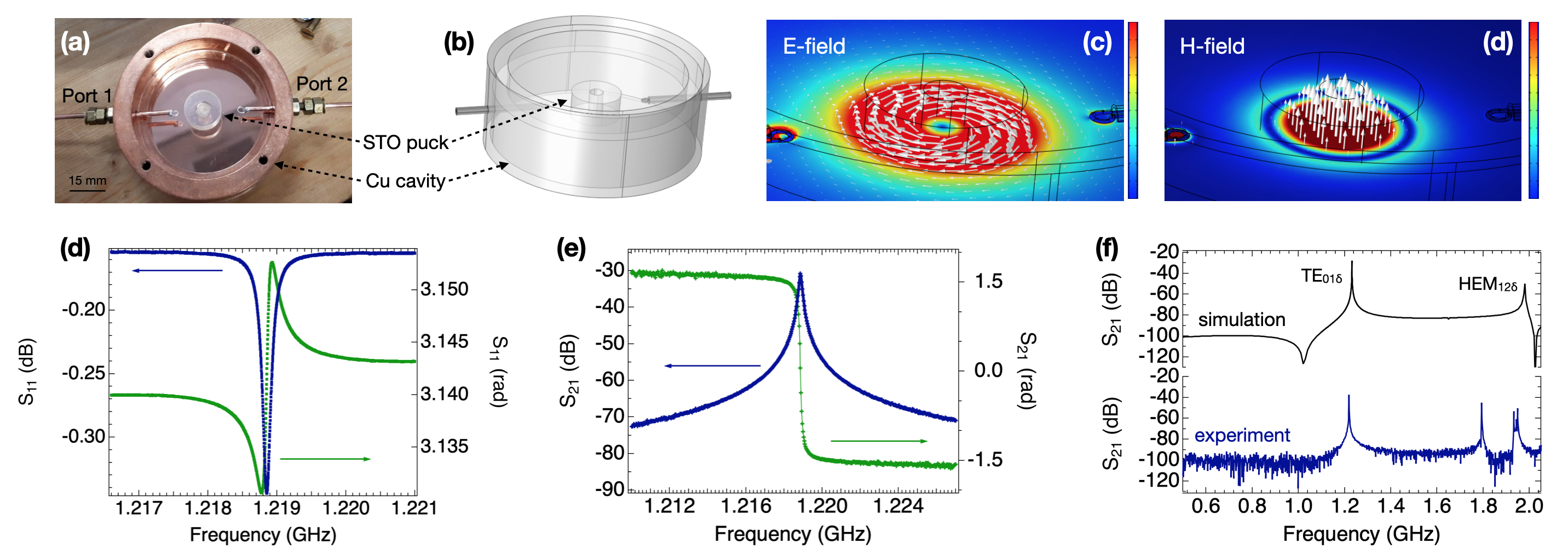}
		\caption{Characterization of the STO resonator at room temperature. (a) Photograph showing the STO puck and Cu cavity used in the experiments. (b) Sketch of the model used for simulations (COMSOL Multiphysics). The top cap is not shown. (c,d) Simulated distribution of the root-mean-square electric and magnetic field for the TE$_{01\delta}$ mode ($\varepsilon_r=318$). (d,e) Plots of reflection ($S_{11}$) and transmission spectra ($S_{21}$) measured at room temperature (incident power 0~dBm). The amplitude is shown in blue, the phase in green. (f) Comparison between simulated and experimental $S_{21}$ spectra. The peak at \unit[1.22]{GHz} is reproduced by the the simulated $\mathrm{TE_{01\delta}}$ mode, while the dip displayed by the simulation at $\approx \unit[1]{GHz}$ is below the background transmission due to the noise floor of the measurement ($\unit[-100]{dB}$). The peak at \unit[1.8]{GHz} is probably related to the hybrid $\mathrm{HEM_{12\delta}}$ mode, although in this case the simulation shows a mismatch of $\approx \unit[200]{MHz}$.}
		\label{fig:STO_room_temp}
    \end{figure*}

The use of resonant cavities for sensitive measurements has a rich history, particularly in the fields of fundamental physics, metrology, and quantum mechanics. For example, early work by Pound and Rebka (1959) demonstrated the potential of resonant cavities to detect extremely small frequency shifts, paving the way for their use in precision measurement \cite{Pound59}. 
In the field of particle physics, high-$Q$ microwave cavities have been instrumental in the search for axions, hypothetical particles proposed as dark matter candidates \cite{Asztalos11, Du18}. Similarly, resonant cavities are used in experiments testing the fundamental constants of nature, providing critical tests of physical theories beyond the Standard Model \cite{Parker18, Nagel15, Lo16}. In quantum information science, superconducting microwave cavities are employed to couple qubits and facilitate the quantum state readout with high fidelity \cite{Blais04, Reagor16}. Furthermore, hybrid quantum systems combining resonant cavities with other quantum technologies, such as optomechanical or spin systems such as nitrogen-vacancy centers in diamond, have shown promise for enhanced sensitivity in detecting weak forces and fields. These hybrid systems leverage the strong coupling between cavity modes and quantum states to achieve unprecedented measurement sensitivity \cite{Barzanjeh16}. The inherent sensitivity of resonant cavities also extends to applications in precision spectroscopy and time-keeping \cite{Bloom14, Matei17, Newbury11}.

The coupling of two or more resonators has been exploited for several of the applications mentioned above \cite{GallopIEEE01, Axline18, Chou24, Majumdar12, MettRSI08, ElnaggarJMR14_1, ElnaggarJMR14_2, Wang17, Bonizzoni18}. If a high $Q$ factor resonator provides a stable reference frequency, the coupling with a second resonator can be used as an additional resource and to make the system sensitive to a specific physical quantity. For example, superconducting cavities were coupled in different ways, showing that the photon lifetime of the local field can exceed that of the bare cavities~\cite{Wang24}. Coupled sapphire resonators showed an increase in the $Q$ factor and tunability due to the supermode effect, which was studied for axion detection with high sensitivity \cite{McAllister18}. Moreover, systems composed of coupled dielectric resonators were investigated for the bolometric detection of particles and X-ray photons \cite{GallopIEEE01, Hao05, Hao20, Hao21}. There, a sapphire resonator that acts as a stable reference frequency is coupled to a second resonator that has a strongly temperature-dependent permittivity (e.g.~CaTiO$_3$ or SrTiO$_3$). Absorption of energy causes a change in the permittivity of the absorber, which in turn produces a measurable change in the resonant frequency of the coupled system. Additionally, coupled resonator systems typically show a transmission zero, or notch, at the frequency for which the injected power is completely reflected back. The frequency and depth of the notch are sensitive probes of the frequency and coupling of the resonant modes \cite{HuanIEEE07, Su17, Bonizzoni18}.

Dielectric resonators have attracted long-time interest for several applications in sensing, metrology, and detectors due to their well-defined resonant modes and high $Q$ factors. High permittivity materials such as SrTiO$_3$, KTaO$_3$, LaAlO$_3$ and TiO$_2$ have been used in different situations, including room temperature masers \cite{Breeze15}, dark matter searches \cite{Brun19} quantum sensing \cite{Eisenach21} and spin manipulation \cite{Vallabhapurapu21}. In particular, strontium titanate (SrTiO$_3$) has been the subject of intense study due to its peculiar physics \cite{PaiRepProgPhs18}, characterized in particular by quantum paraelectricity \cite{CowleyPR64, Muller79, Viana94} and superconductivity \cite{SchooleyPRL64} at low temperature, multiferroicity \cite{WangNatCommun24} and emergent conductivity or magnetism at the interface with other oxides, i.e.~in LaAlO$_3$/SrTiO$_3$ heterostructures \cite{PaiRepProgPhs18}. The properties of SrTiO$_3$ microwave resonators were also investigated \cite{KrupkaIEEE94, GeyerJAP05, HosainJAP19, ZhaoIEEE22}, showing permittivity values reaching 30000 at 10~K and recently reporting an anomalous behavior below 1~K \cite{DavidovikjPRB17, Engl19}.

Here we report the study of a particular system of coupled resonators, in which a strontium titanate (hereafter STO) puck, whose resonant frequency is strongly dependent on temperature or other external parameters, is placed in a three-dimensional high-$Q$ cavity. In particular, we consider as a prototype an elliptical TESLA-shaped cavity with fundamental mode at 1.3~GHz, whose geometry has been developed in the context of particle accelerators and optimized for manufacture with superconducting materials to obtain $Q$ factors as high as $10^9$ \cite{TESLA00, PadamseeWiley23}, with potential applications also in dark matter searches \cite{RomanenkoPRL23} and quantum computing \cite{RoyProcFermilab23}. We first characterize the STO resonator alone and test the evolution of permittivity and losses at temperatures down to 160~mK. We then exploited finite element electromagnetic simulations to study the coupling between the elliptical cavity and the dielectric resonator, discussing the tuning of frequency and $Q$ factor of the coupled modes as a function of geometric and material parameters, such as the position of the STO puck within the cavity and the permittivity of the STO puck. These results demonstrate great tunability for the proposed coupled system and allow the definition of conditions to control the coupling between high-$Q$ cavities and high-permittivity dielectric resonators, opening a path for the exploitation of these systems in applications. In particular, we discuss the use of the coupled cavity-STO resonator system in a bolometer operating at different temperatures.

		\section{Characterization of the STO resonator}	\label{sec:STO_characterization}

The dielectric resonator was realized with a STO crystal produced by SurfaceNet GmbH that was shaped into a cylinder having radius $a$=8.17~mm and height $d$=7.26~mm, with a central hole of radius 2~mm. For the characterization of the dielectric resonator we used a hollow copper cavity (inner radius 60~mm, height 22~mm), whose lowest mode at 3.8~GHz stands well above the fundamental mode of the dielectric resonator in order to avoid cross-couplings between their modes. The STO puck was installed on top of two quartz spacers and coupled to the coaxial lines by loop antennae positioned at a distance of 10-20 mm from the outer surface of the cylinder (Fig.~\ref{fig:STO_room_temp}(a)). Measurements were carried out with the whole set-up installed at room temperature, in liquid nitrogen, or in a dilution fridge.

    \begin{figure}
		\centering
        \includegraphics[width=0.9\columnwidth]{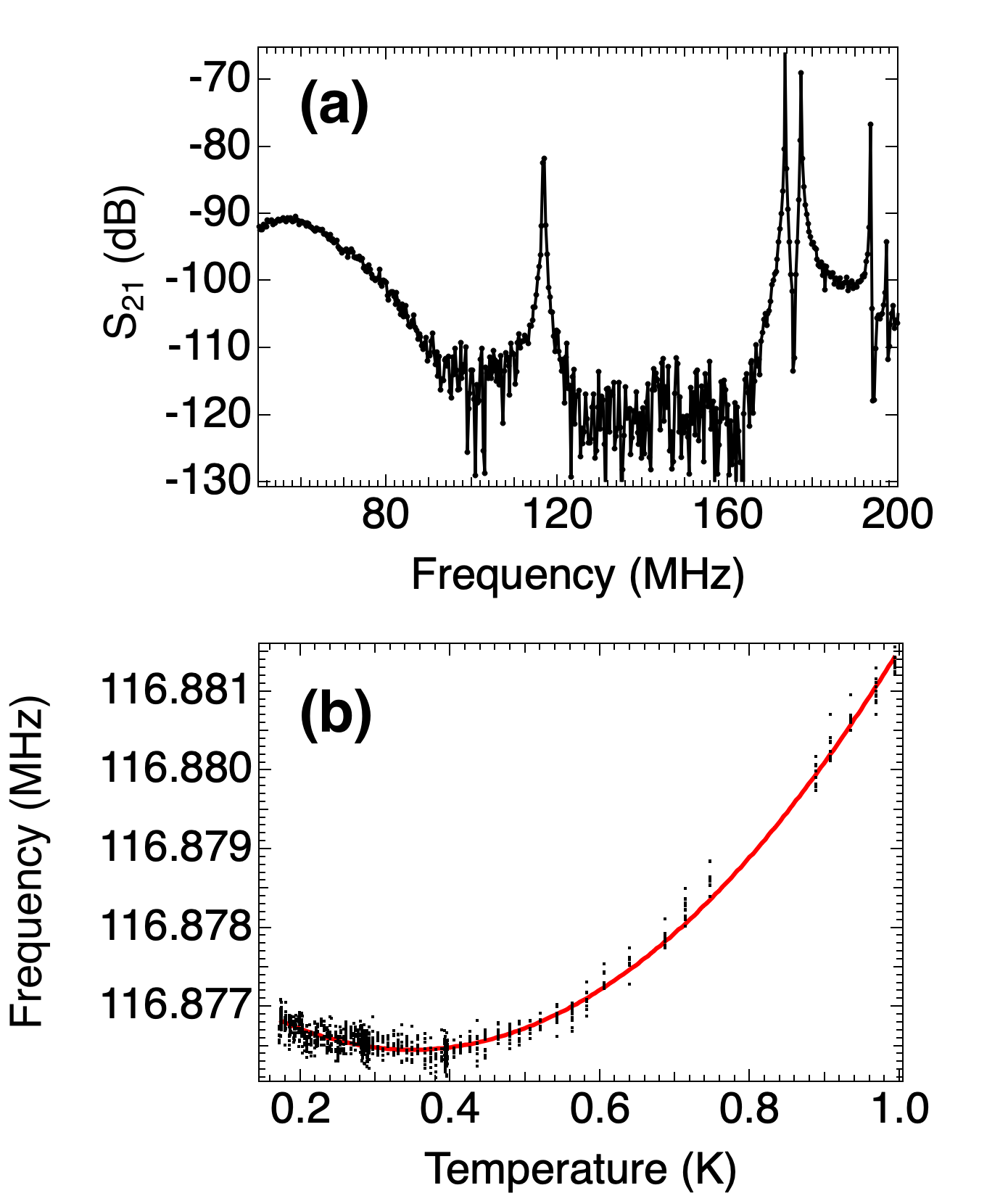}
		\caption{Characterization of the STO resonator at temperature below 1~K. (a) Transmission spectrum acquired at 170~mK. (b) Temperature dependence of the fundamental mode frequency. The red solid line shows the fit of the experimental points}.
		\label{fig:STO_low_temp}
    \end{figure}

\paragraph*{Room temperature measurements.} Reflection and transmission spectra acquired at room temperature show the fundamental mode of the STO resonator at frequency $1.22$~GHz (Fig.~\ref{fig:STO_room_temp}(d,e)). Considering the geometric factors of the resonator and the STO permittivity $\epsilon_r \approx 318$, we find that this number corresponds well to the frequency of the $\mathrm{TE_{01\delta}}$ mode derived by the semi-analytical formula \cite{kajfez98}
\begin{equation} \label{AnalyticFormula}
    f_\mathrm{STO} \approx\frac{\alpha}{\sqrt{\varepsilon_\mathrm{r}}}\
\end{equation}
being $\alpha \mathrm{[GHz]}=34/(a\mathrm{[mm]})\left(a/d+3.45\right)$, while $a$ and $d$ are the radius and height of the puck, respectively. The loaded quality factor extracted from the transmission spectrum (Fig.~\ref{fig:STO_room_temp}(e)) is $Q_L=8000$. Since the loops are located far apart from the dielectric puck, the small coupling between the feedlines and the resonator results in an insertion loss of $\unit[30]{dB}$. Thus, the internal quality factor can be approximated as $Q_0\approx Q_L\approx 1/\tan{\delta}$. The loss tangent of the STO results $\tan{\delta}=1.2 \times 10^4$. Finite element electromagnetic simulations confirm the expected distribution for the $\mathrm{TE_{01\delta}}$ mode (Fig.~\ref{fig:STO_room_temp}(c,d)), whose frequency results about 1.2~GHz in close agreement with the experimental data (Fig.~\ref{fig:STO_room_temp}(f)).

\paragraph*{Low temperature behavior.} The temperature evolution between 4 and 300~K of cylindrical STO resonators is reported in the literature \cite{KrupkaIEEE94, GeyerJAP05, ZhaoIEEE22, HosainJAP19}. In this temperature range, the maximum value of the $Q$ factor of 20000 is reached around 100~K, just above the structural temperature transition of the STO. The nonlinear change in the permittivity of STO encompasses two orders of magnitude to obtain $\varepsilon_r \approx 3 \times 10^4$ and $\tan{\delta} \approx 1 \times 10^{-4}$ at 5~K \cite{GeyerJAP05}.

We tested the STO resonator at temperatures below 1~K (Fig.~\ref{fig:STO_low_temp}). The STO resonator was slowly cooled to 160 mK in about 6~hours. The temperature was measured using a thermometer installed in the copper cavity. At the lowest temperature, the frequency of the $\mathrm{TE_{01\delta}}$ mode resulted in $116.9$~MHz with $Q \approx 1 \times 10^4$; these values suggest $\varepsilon_r \approx 3 \times 10^4$ and $\tan{\delta} \approx 1 \times 10^{-4}$. As a function of the decreasing temperature, the frequency of the $\mathrm{TE_{01\delta}}$ mode decreases until 0.4~K, while for lower temperatures we observe a slight upturn. In this temperature ($T$) range, the frequency of the resonator can be reproduced using a polynomial curve, $f_{STO}=K_0+K_1 T+K_2 T^2$. The fit of the experimental points (Fig.~\ref{fig:STO_low_temp}(b)) gives $K_0=116.88~\unit{MHz}$, $K_1=-0.008~\unit{MHz/K}$ and $K_2=0.012~\unit{MHz/K^2}$.

\begin{figure}[!tbh]
		\centering
        \subfloat[]{\includegraphics[width=0.63\columnwidth]{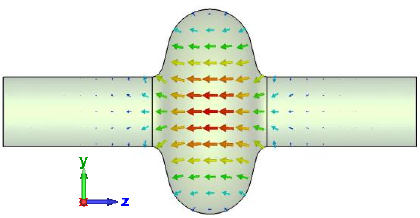}}\hfill
        \subfloat[]{\includegraphics[width=0.37\columnwidth]{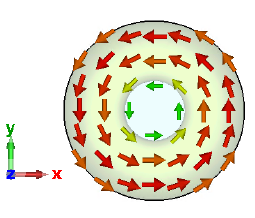}}\\
        \subfloat[]{\includegraphics[width=0.5\columnwidth]{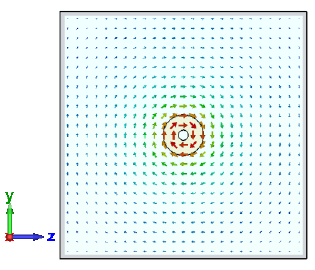}}\hfill
        \subfloat[]{\includegraphics[width=0.5\columnwidth]{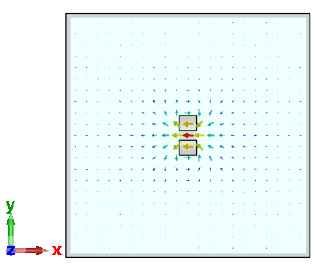}}\\
		\caption{Field distributions for the $\mathrm{TM}_{010}$ mode in a TESLA-shaped \unit[1.3]{GHz} accelerating cavity: (a) electric field and (b) magnetic field. Additionally, field distributions for the first mode of the STO puck: (c) electric field and (d) magnetic field.}
	\label{fig:FMFIelds}
\end{figure}

	\begin{figure}
		\centering
        \subfloat[]{\includegraphics[width=0.5\columnwidth]{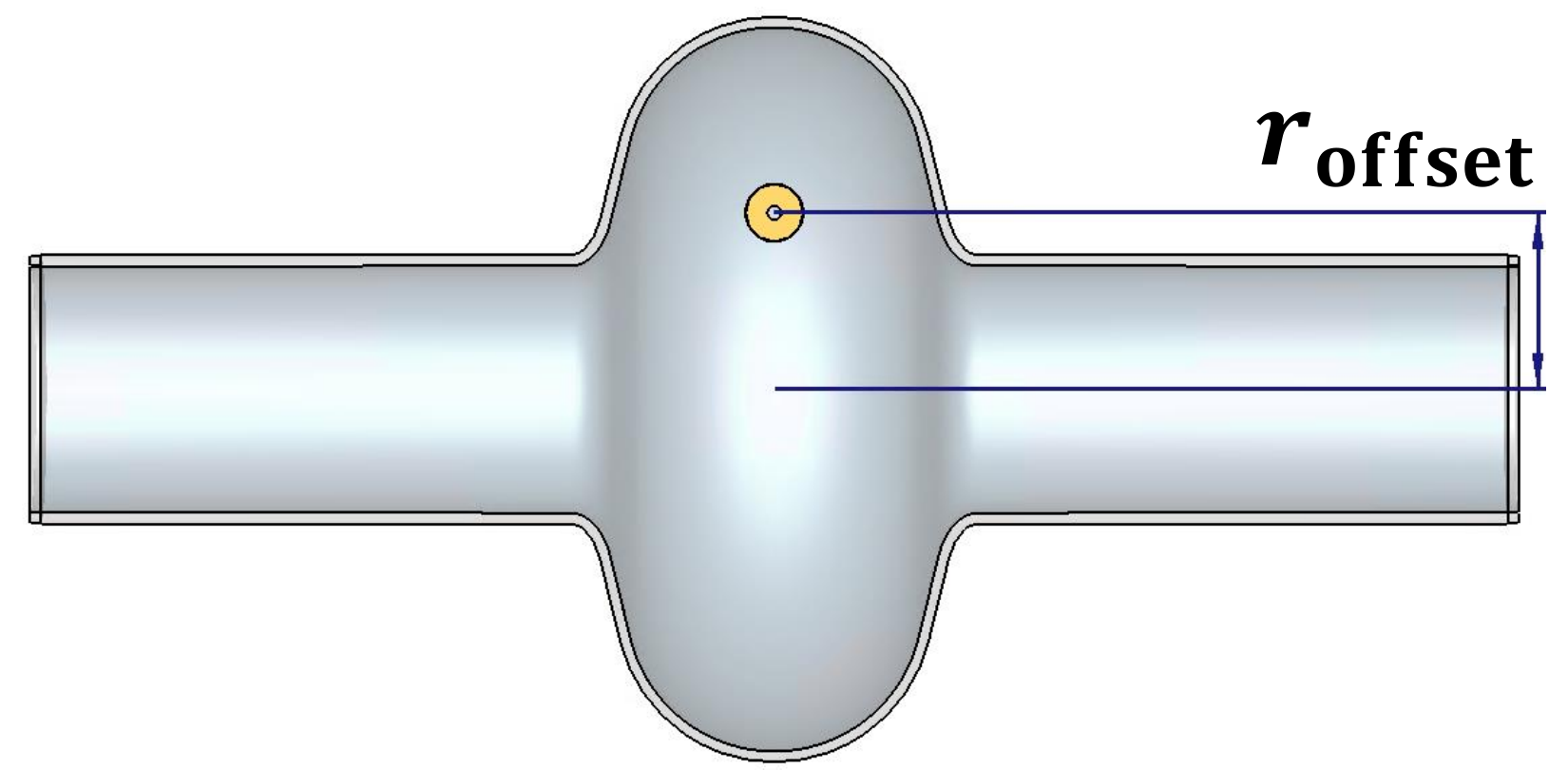}}\hfill
        \subfloat[]{\includegraphics[width=0.5\columnwidth]{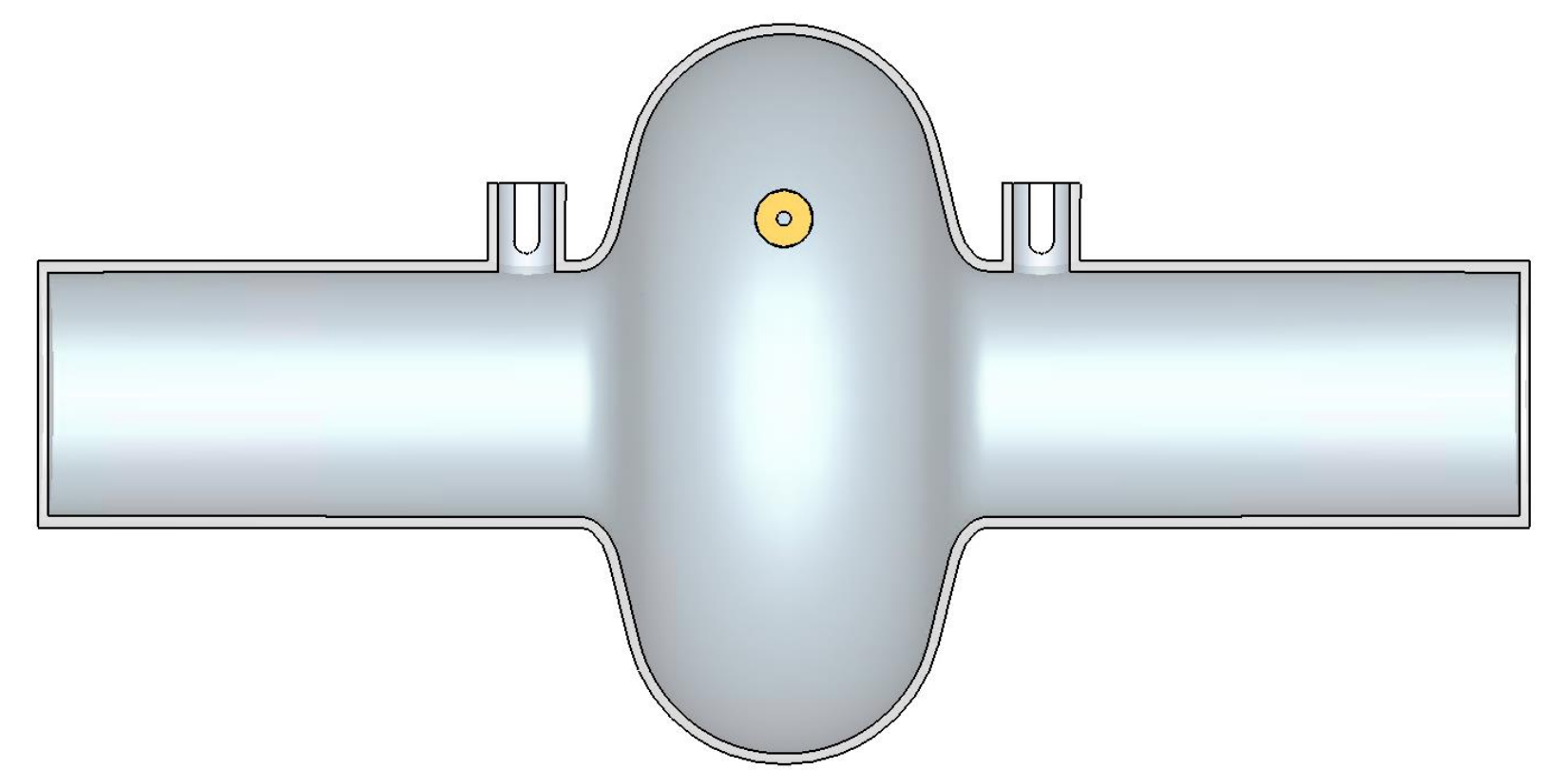}}\\
		\caption{Simulation setups of the coupled elliptical cavity and STO puck. The radial offset of the STO puck ($r_\mathrm{offset}$) influences the coupling strength between the two modes. (a) Eigenmode analysis setup, and (b) S-parameter analysis setup. }
		\label{fig:SimulationSetup}
	\end{figure}

		\begin{figure}[htp!]
		\centering
        \subfloat[]{\includegraphics[width=1\columnwidth]{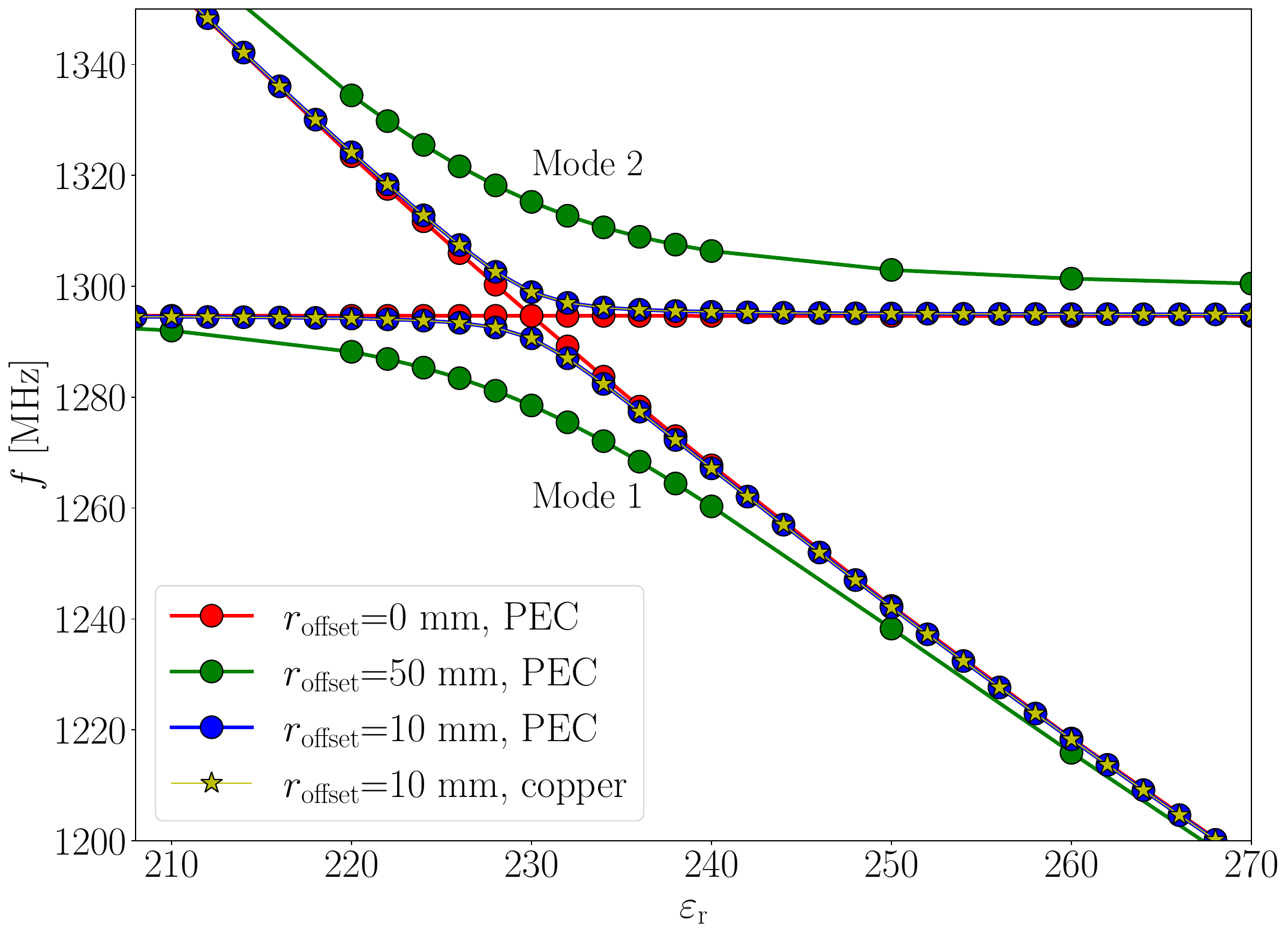}}\hfill
        \subfloat[]{\includegraphics[width=1\columnwidth]{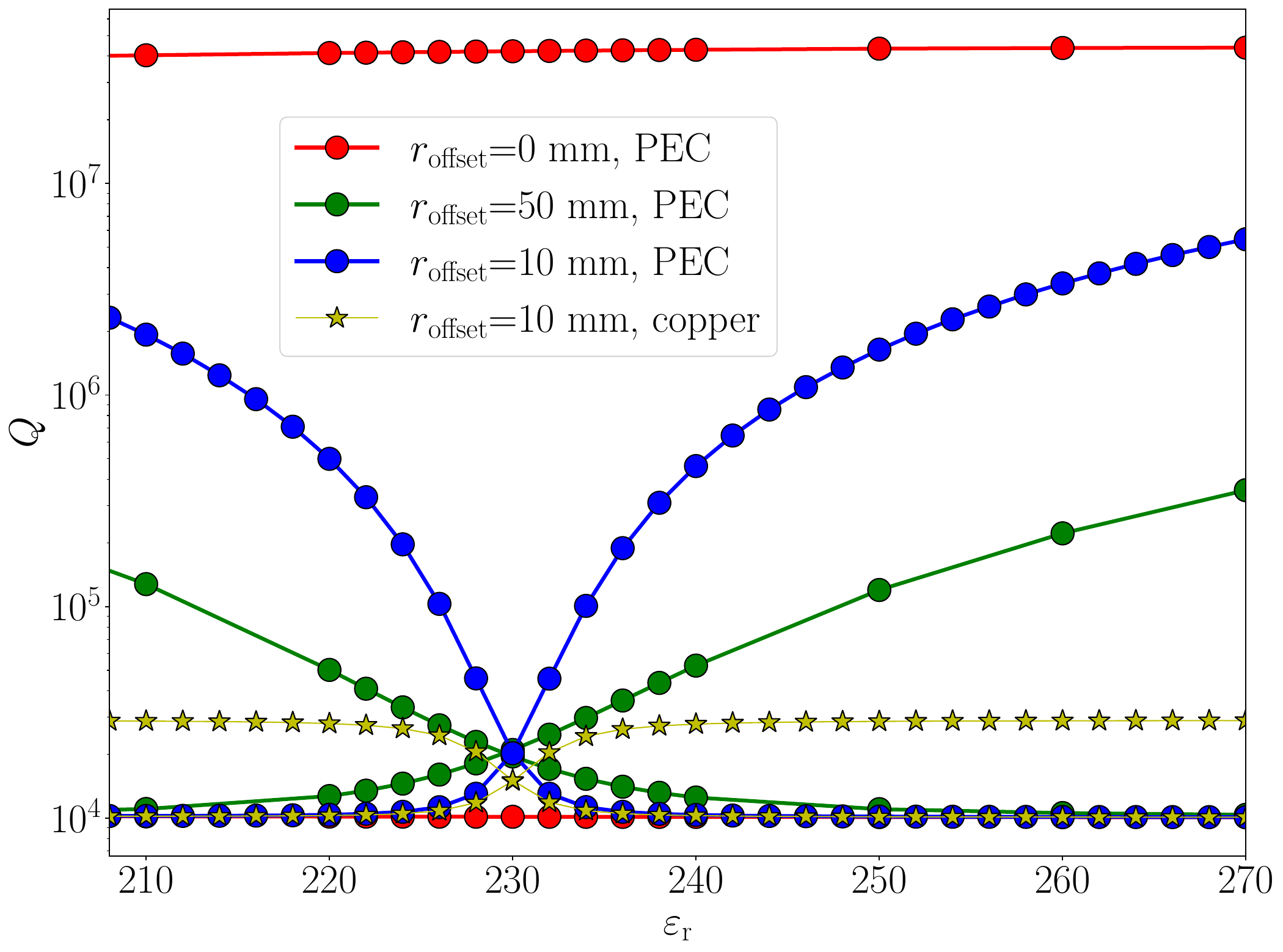}}\hfill
         \subfloat[]{\includegraphics[width=1\columnwidth]{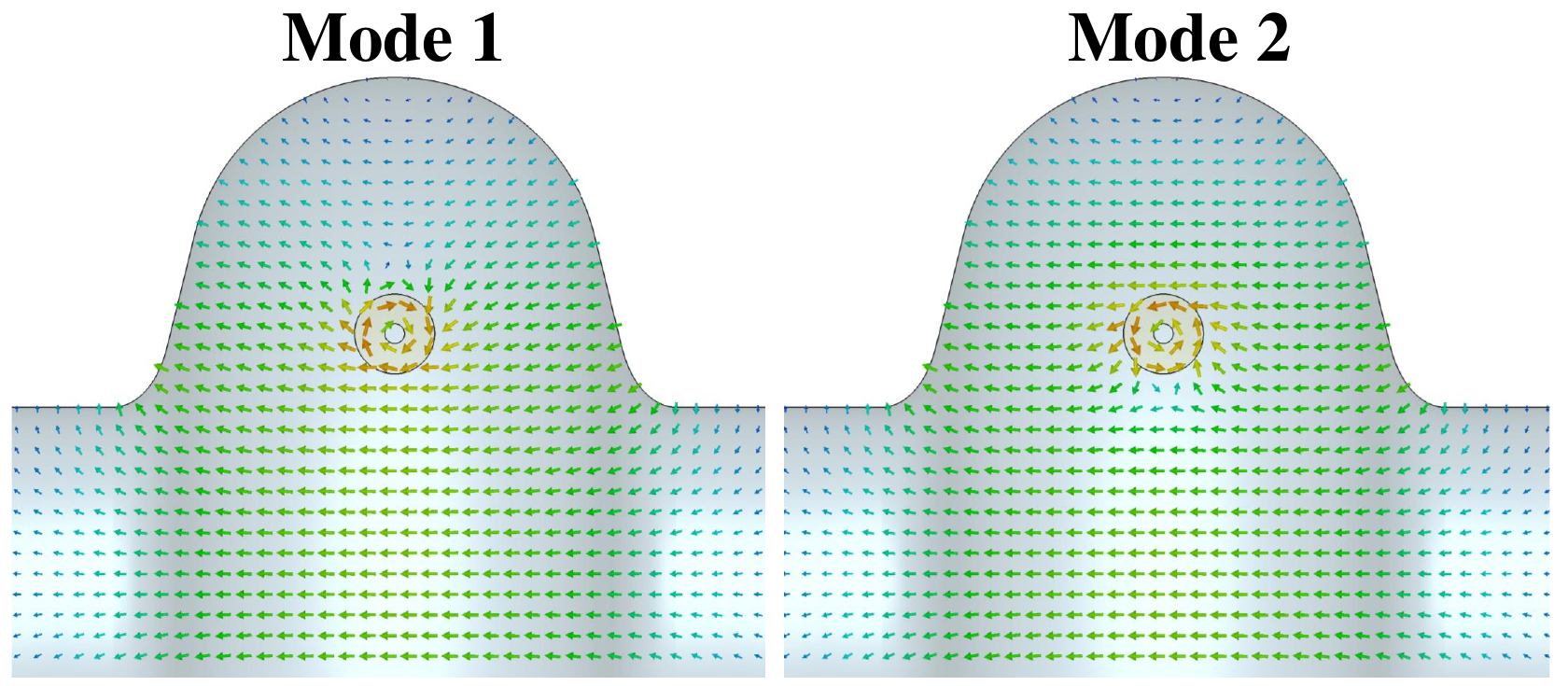}}\\
		\caption{Parameter sweep of the dielectric constant ($\varepsilon_r$) of STO and its effect on the frequency and quality factor of the first two modes of the coupled elliptical cavity and STO puck at different radial offset ($r_\mathrm{offset}$) values. Mode 1 is the mode with the lower frequency. (a) Frequency of the eigenmodes, (b) quality factor of the eigenmodes, and (c) E-field distributions of the two modes at $\varepsilon_\mathrm{r}=230$ and $r_\mathrm{offset}=50$~mm.  }
		\label{fig:CouplingEigenmode}
	\end{figure}

 		\begin{figure}[htp!]
		\centering
        \subfloat[]{\includegraphics[width=0.5\columnwidth]{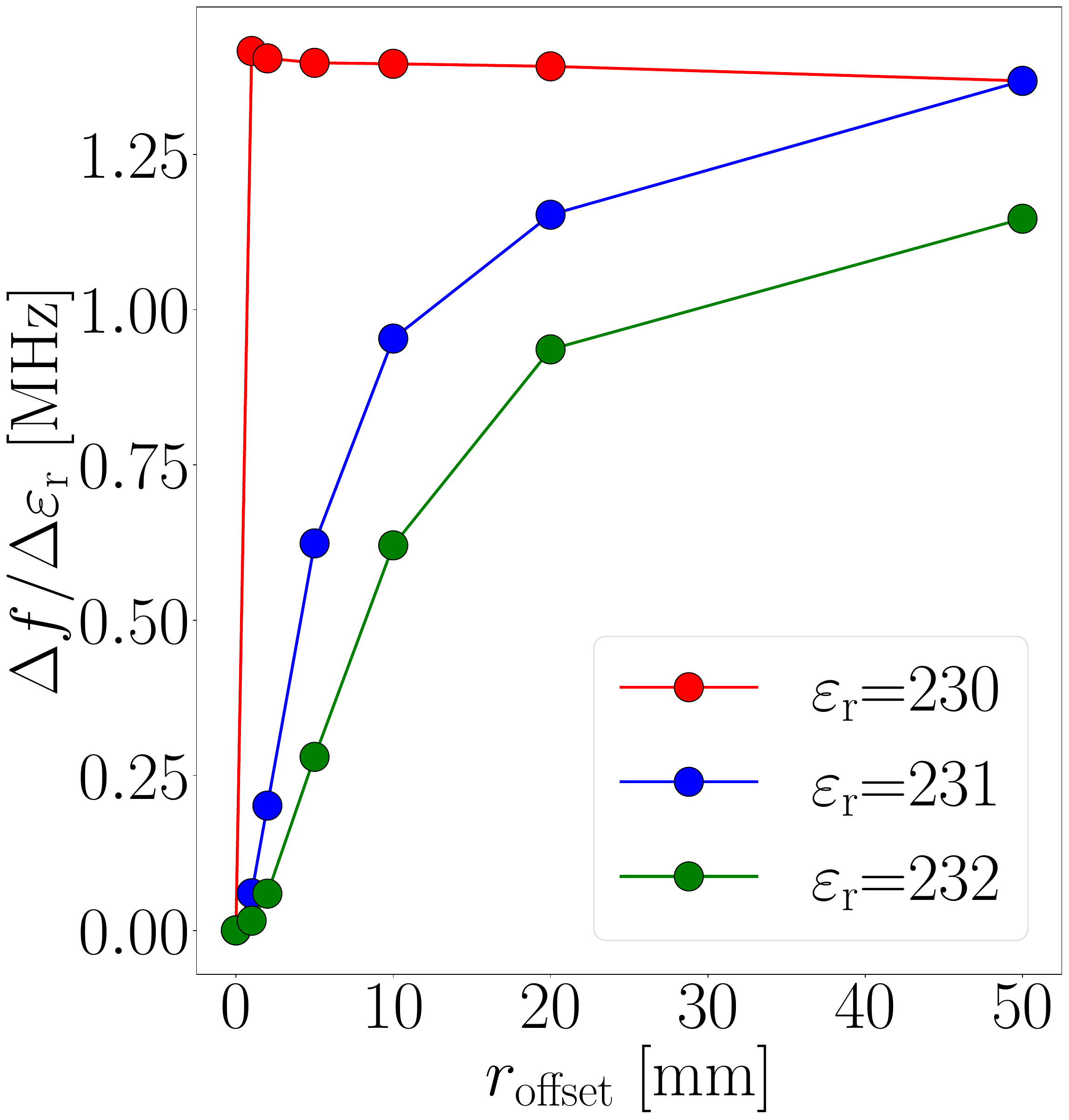}}\hfill
        \subfloat[]{\includegraphics[width=0.5\columnwidth]{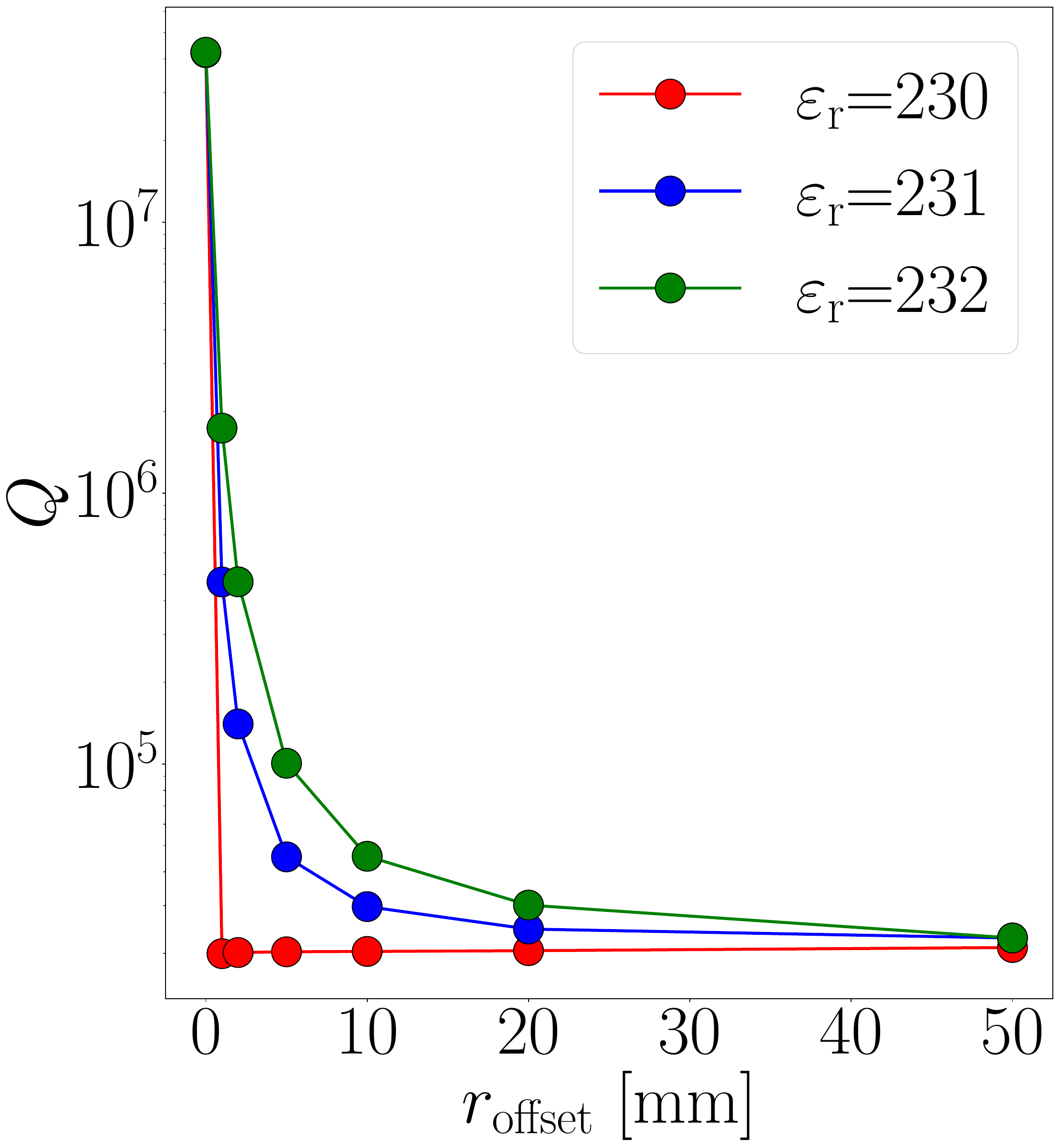}}\hfill
		\caption{Dependency of the frequency sensitivity (a) and quality factor (b) of the elliptical cavity-dominant mode with respect to $r_\mathrm{offset}$ at different values of $\varepsilon_r$.}
		\label{fig:OffsetSensitivity}
	\end{figure}

		\begin{figure}[h!]
		\centering
        \subfloat[]{\includegraphics[width=1\columnwidth]{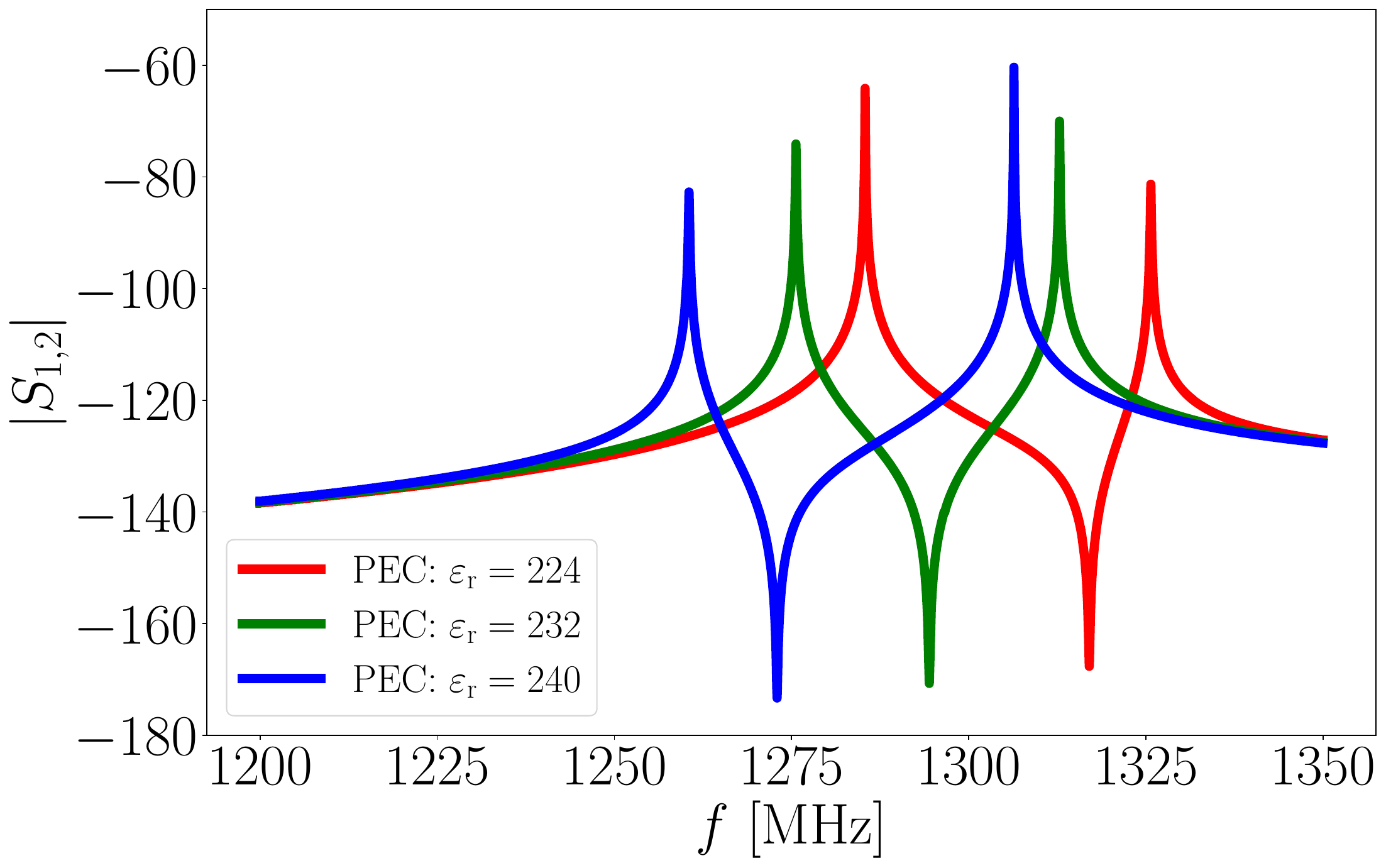}}\hfill
        \subfloat[]{\includegraphics[width=1\columnwidth]{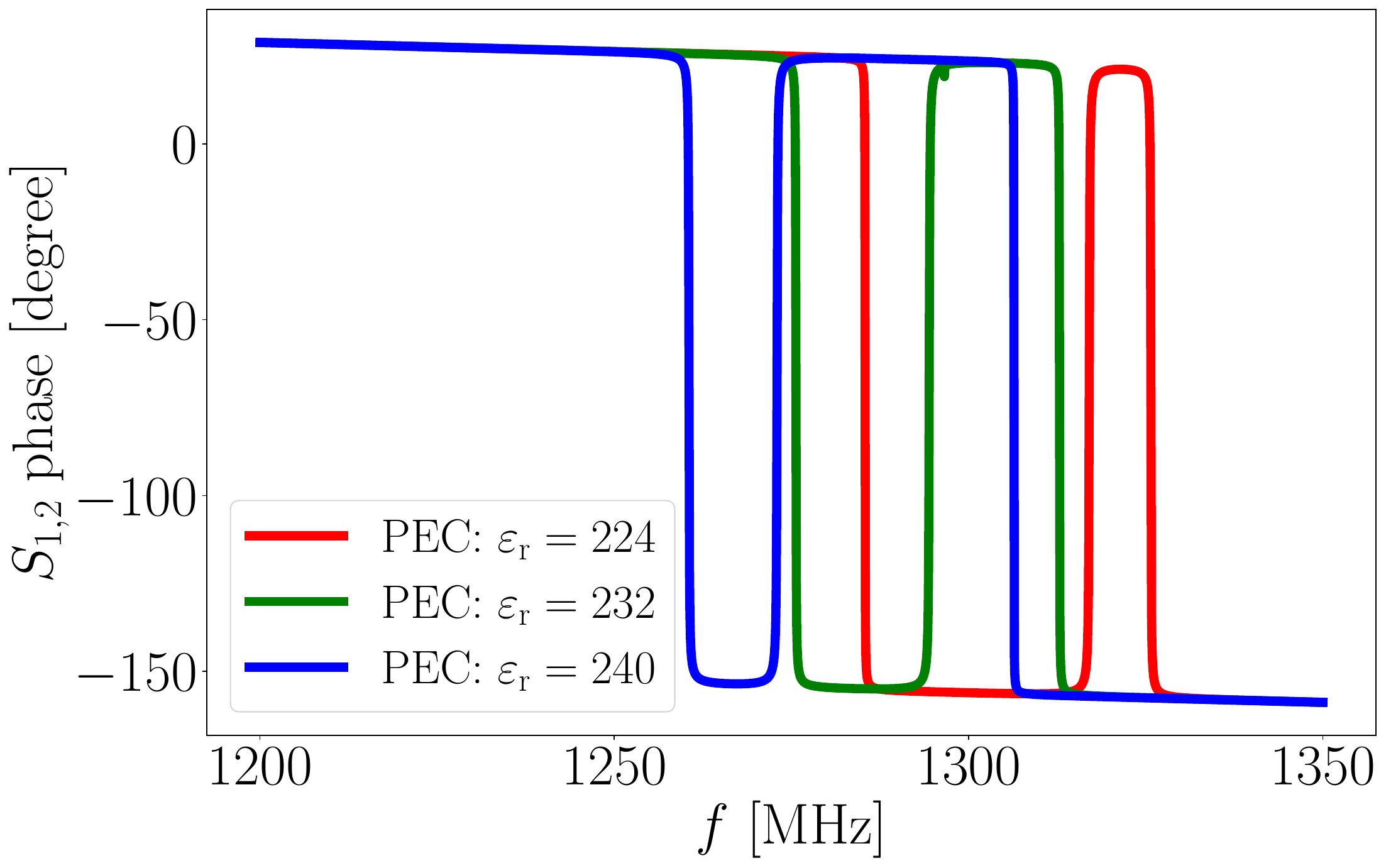}}\hfill
        \subfloat[]{\includegraphics[width=1\columnwidth]{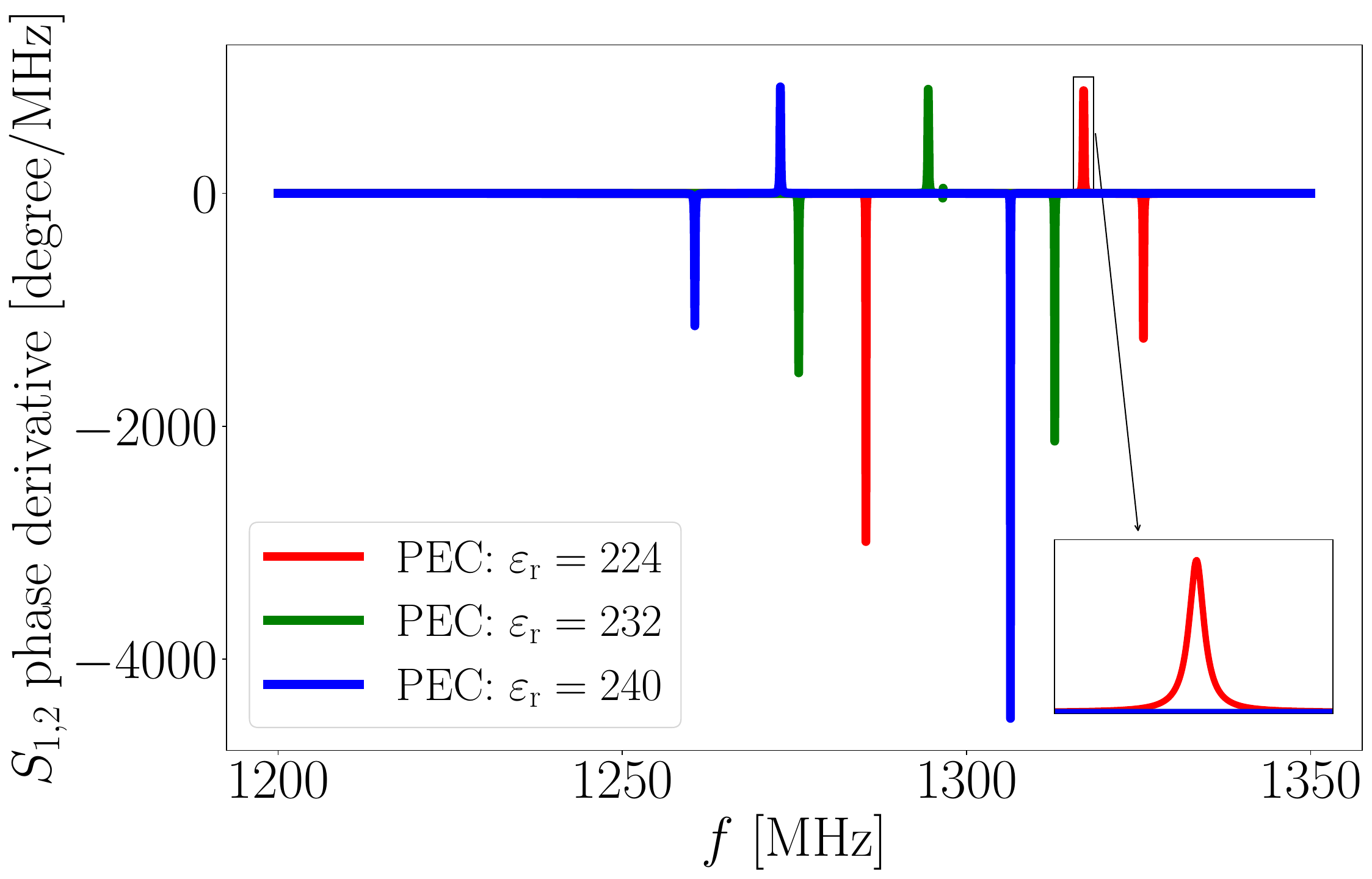}}\hfill\\
		\caption{S-parameters of the excitation scheme shown in Fig.~\ref{fig:SimulationSetup} at different $\varepsilon_r$ and $r_\mathrm{offset}=\unit[50]{mm}$ for a PEC cavity. (a) Magnitude of $S_{1,2}$, (b) Phase ($\varphi$) of $S_{1,2}$, and (c) Phase derivative ($\mathrm{d}\varphi/\mathrm{d}f$) of $S_{1,2}$.  
        Varying $\varepsilon_r$  shifts the resonance frequencies of the two peaks and the notch. The phase derivative is linked to the quality factor, with higher amplitude for the mode with the larger Q-factor, and shows minimal change at the notch as $\varepsilon_r$ varies.}
		\label{fig:CouplingSParameters}
	\end{figure}

	\begin{figure*}[htp]
		\centering
        \subfloat[]{\includegraphics[width=0.85\columnwidth]{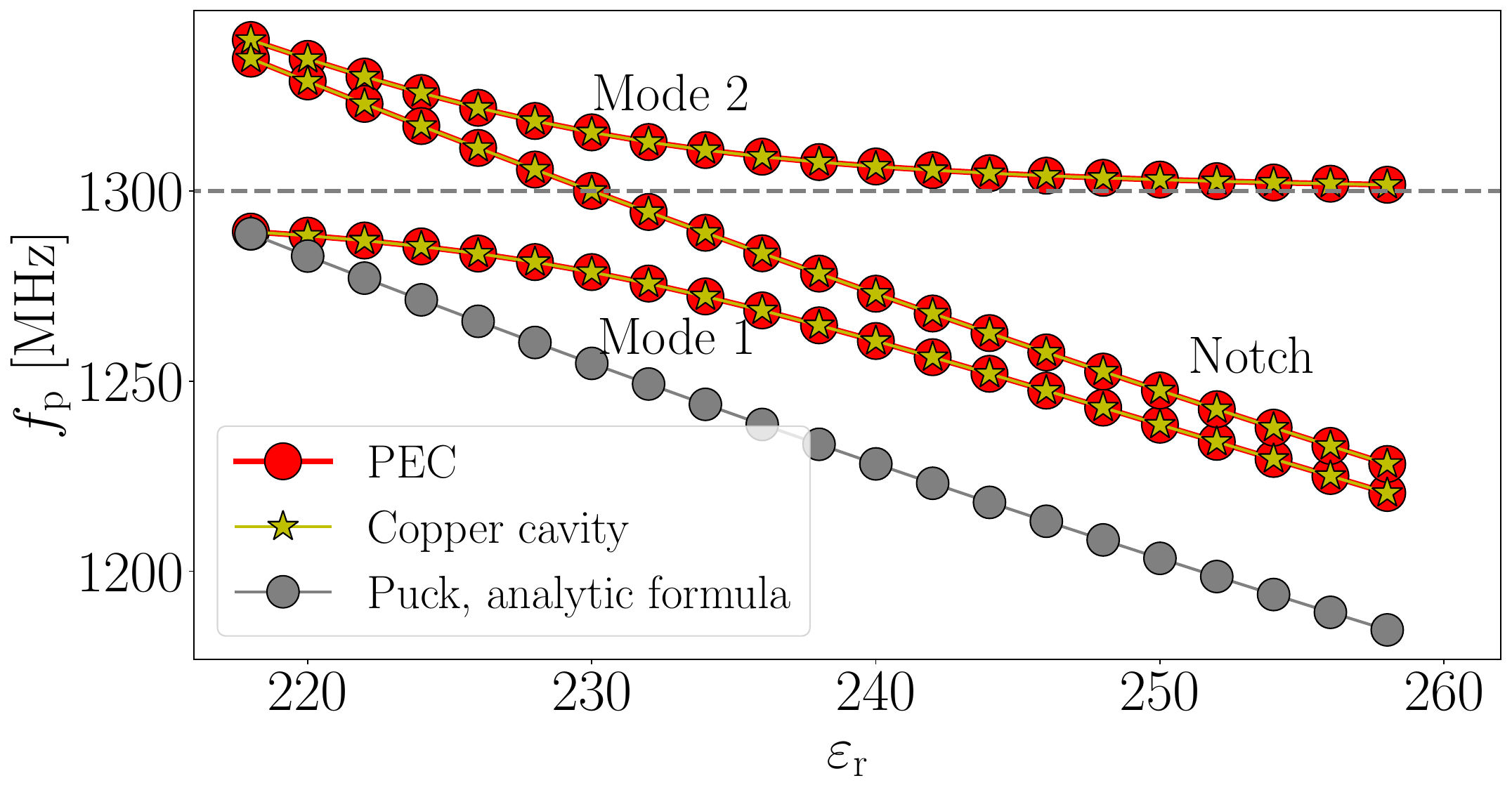}}\hfill
        \subfloat[]{\includegraphics[width=0.84\columnwidth]{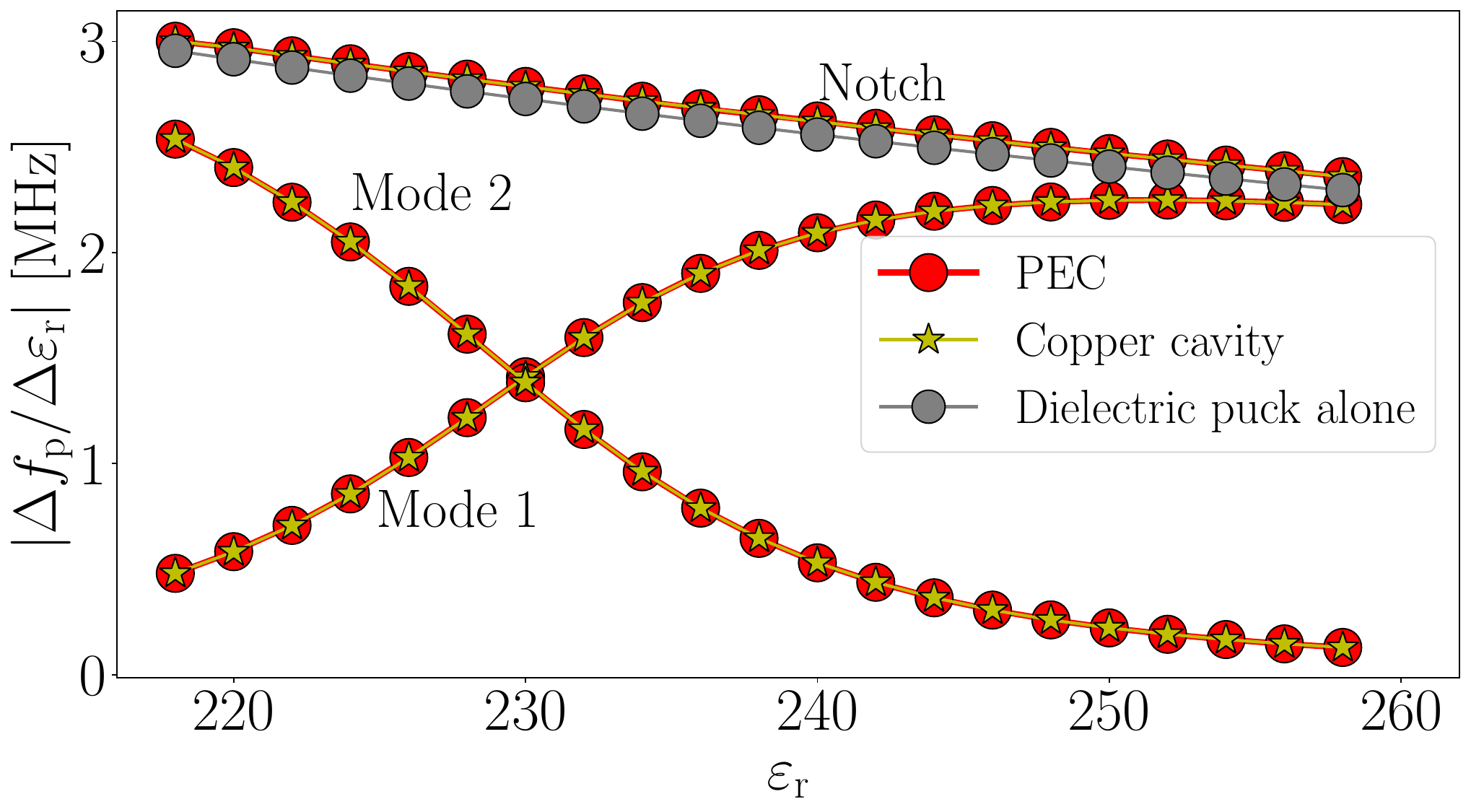}}\hfill
         \subfloat[]{\includegraphics[width=0.85\columnwidth]{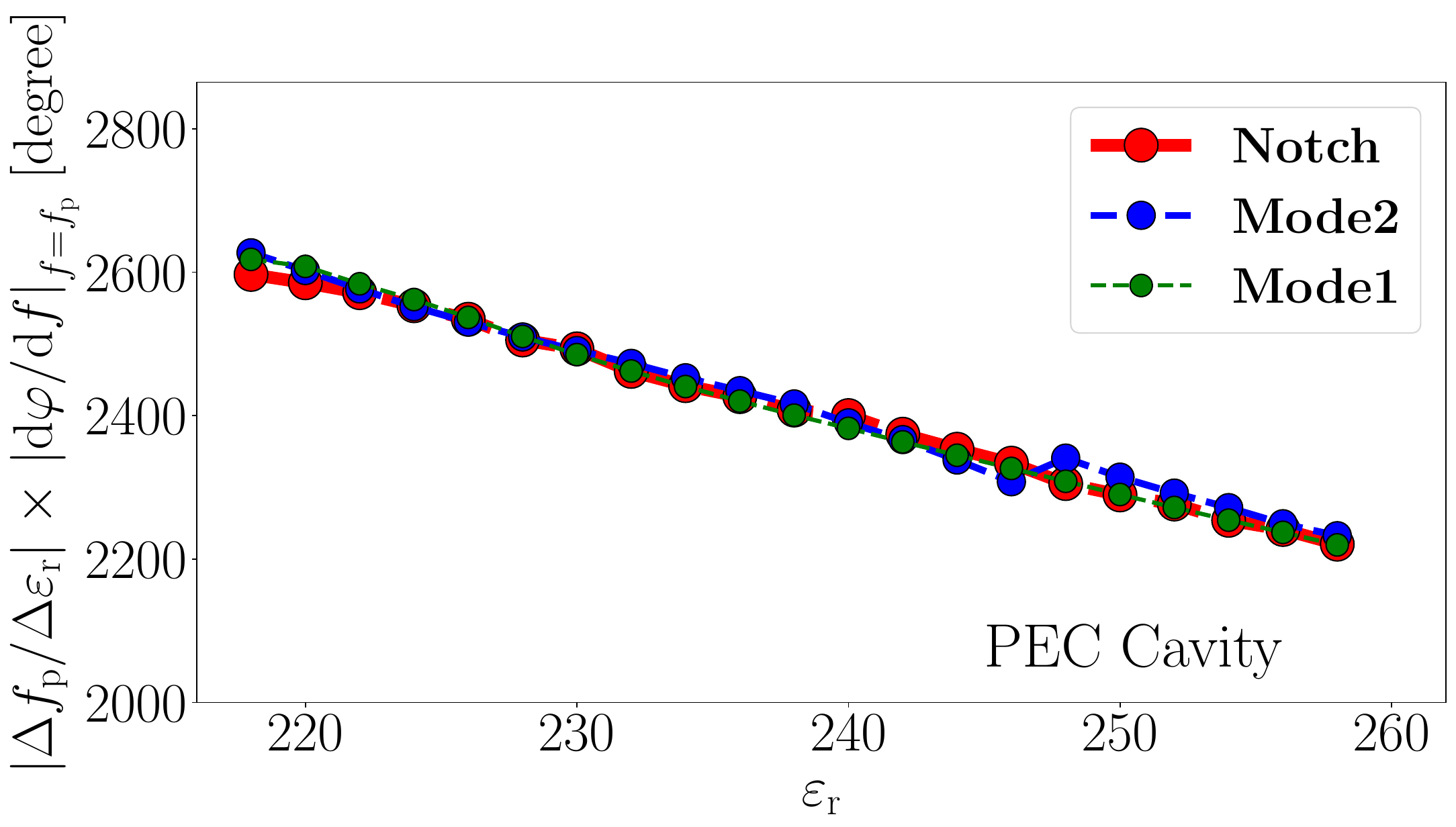}}\hfill
        \subfloat[]{\includegraphics[width=0.85\columnwidth]{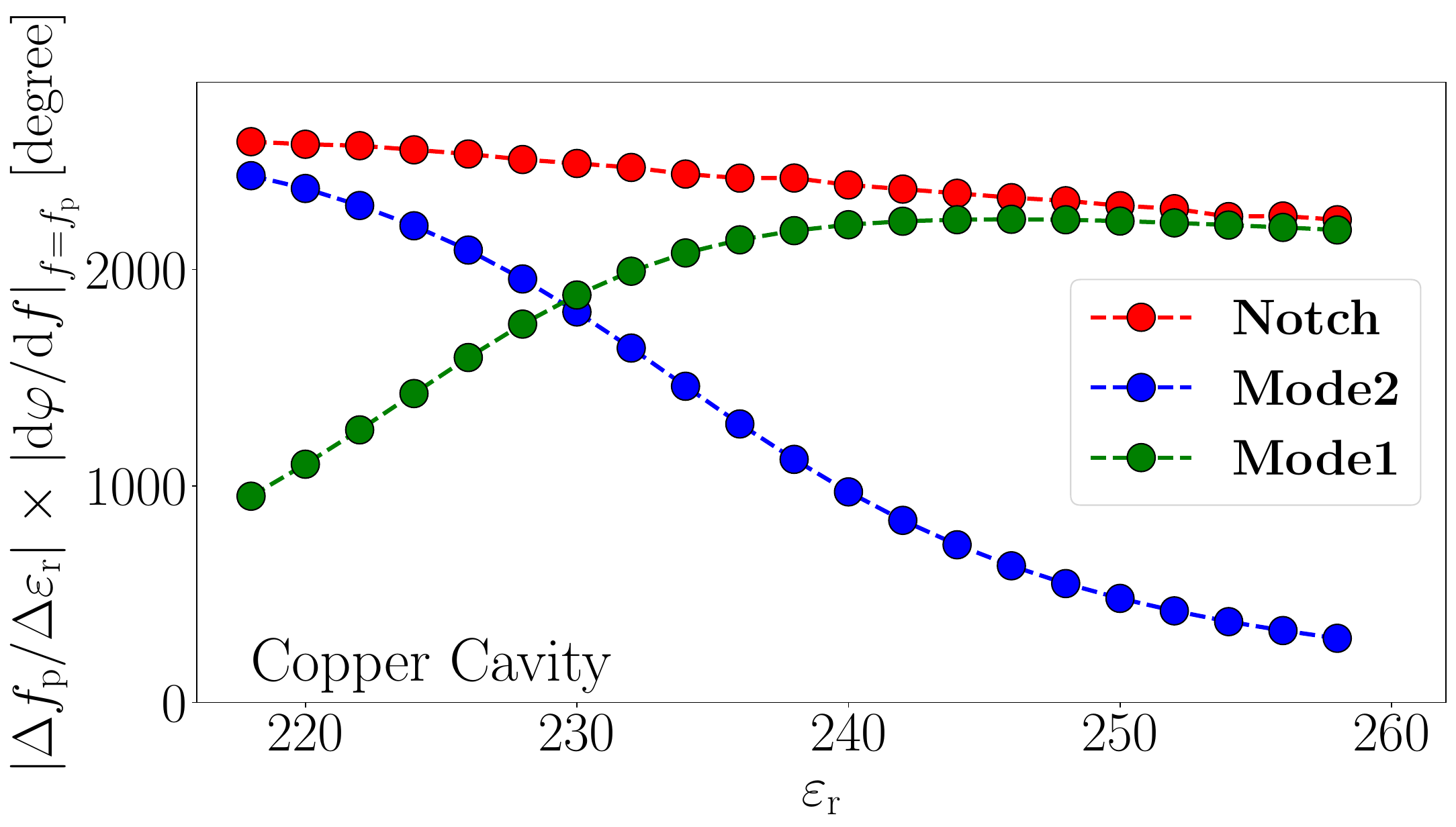}} \\
		\caption{Analysis of the S-parameters shown in Fig.~\ref{fig:CouplingSParameters}. 
        (a) Frequencies of the first and second peaks, as well as the notch, of the $|S_{1,2}|$ curves shown in Fig.~\ref{fig:CouplingSParameters}(a). (b) Derivative of the  curves from (a). (c) and (d) show the curves in (b) multiplied by the phase derivative at $f = f_\mathrm{p}$ (from Fig.~\ref{fig:CouplingSParameters}(c)), for a PEC cavity and a copper cavity, respectively. 
        Mode 1 and Mode 2 correspond to the frequencies of the first and second peaks of the $|S_{1,2}|$ curves, respectively, and the notch corresponds to the frequency of the dip between these two peaks. The frequency of the puck alone and its derivative calculated by the semi-analytic formula is also shown in (a) and (b).   The product of frequency sensitivity and the maximum phase derivative is almost equal at the notch for both the PEC and copper cavities (red curves in (c) and (d)). However, for modes 1 and 2, it is smaller in the copper cavity compared to the PEC cavity near $\varepsilon_\mathrm{r}$ of 230, where the coupling is at its maximum.}
		\label{fig:CouplingAntennaDerivatives}
	\end{figure*}

		\section{Coupled STO and elliptical cavity}	\label{sec:CoupledCavity}

This section examines the coupling of the STO puck with a resonating cavity to form a coupled resonator system. To facilitate this investigation, we selected an elliptical TESLA-shaped cavity\cite{TESLA00} operating at frequency $f_{cav}=$\unit[1.3]{GHz}. This frequency was selected because it closely matches the operational frequency of the available STO puck sample at room temperature and because similar elliptical cavity prototypes were available for measurement at the European Organization for Nuclear Research (CERN). However, it is important to emphasize that the choice of cavity and frequency is not limited to the elliptical cavity at \unit[1.3]{GHz}; other cavity types and frequencies can be used based on specific project requirements. The primary aim of this study is to assess the feasibility of coupling between the STO puck and a superconducting RF cavity.

\paragraph*{Electromagnetic simulations.} The TESLA-shaped elliptical cavity and its fundamental mode (FM), which is a transverse magnetic (TM) mode, are illustrated in Fig.~\ref{fig:FMFIelds}. The radius of the equator of the cavity was adjusted to achieve a frequency of \unit[1.3]{GHz}. For a copper cavity, the quality factor of this mode at room temperature is $2.89\times10^4$. Assuming $\varepsilon_r=318$, the first mode of the STO puck has a frequency of 1.10 GHz and a quality factor of $1.01\times10^4$. The field plot of this mode is also shown in Fig.~\ref{fig:FMFIelds}. The eigenmode and frequency domain simulations in this section were performed using the CST Studio Suite.

To investigate the coupling between the two modes, the STO puck is placed inside the elliptical cavity. Two essential parameters for effective coupling are the orientation of the STO puck within the elliptical cavity and its radial offset from the center. The position of the puck in the elliptical cavity for effective coupling is shown in Fig.~\ref{fig:SimulationSetup}. The coupling is examined using both the eigenmode analysis and the frequency domain analysis, with two identical antennas, each having $Q_\mathrm{ext}=8.6\times10^7$. In the eigenmode analysis, no antennas are considered, and the only losses accounted for are the dielectric losses of STO and surface losses on the elliptical cavity if non-perfect electric conductor (PEC) materials are used.

The results of the eigenmode analysis are presented in Fig.~\ref{fig:CouplingEigenmode}. The investigation examines how the variation $\varepsilon_r$ of the STO, which can be a function of temperature, affects the resonant frequency and the $Q$ factor of the first two modes in the coupled system. The STO-dominant mode is characterized by having most of its energy localized in the STO, while the elliptical cavity-dominant mode retains most of its energy within the elliptical cavity, similar to the uncoupled system. Below $\varepsilon_r=230$, Mode 2 is the STO-dominant mode and Mode 1 is the elliptical cavity-dominant mode. Above $\varepsilon_r=230$, the order of the modes is reversed. 
When $r_\mathrm{offset}=$\unit[0]{mm}, the two modes do not interact as $\varepsilon_r$ changes. In this scenario, the elliptical cavity's frequency (Fig.~\ref{fig:CouplingEigenmode}(a)) and $Q$ factor (Fig.~\ref{fig:CouplingEigenmode}(b)) remain nearly constant at \unit[1.3]{GHz} and $Q_\mathrm{ext}=4.2\times10^7$, respectively, for a PEC cavity. In this PEC situation, the value of the $Q$ factor is influenced by the losses in the dielectric region occupied by the STO puck. In contrast, the frequency of the STO-dominant mode decreases with increasing $\varepsilon_r$, while its $Q$ factor remains nearly constant at $Q_\mathrm{ext}=1.0\times10^4$. In this scenario, the frequencies of the elliptical cavity-dominant mode at \unit[1.3]{GHz} and the STO-dominant mode are not affected by each other and intersect in $\varepsilon_r=230$, without interaction between the modes as $\varepsilon_r$ changes (Fig.~\ref{fig:CouplingEigenmode}(a)). The permittivity value of 230 corresponds to a temperature above room temperature, although the system is ideally intended for operation at lower temperatures where the STO’s resonance frequency exhibits the greatest sensitivity to temperature variations (around 25 K). However, due to the fixed dimensions of the STO puck and the elliptical cavity's frequency, we adjusted the STO permittivity to shift its resonance frequency near 1.3 GHz to investigate the mode coupling between the two resonators. Alternative dielectrics with low-temperature permittivity values closer to those considered in the simulations were also considered. In particular, a rutile puck having similar dimensions shows, at 0.8~K, a mode at 2.6~GHz with $Q$ factor of $5.6 \times 10^5$  \cite{rutile_res}. However, the variation in the permittivity of rutile as a function of temperature is less marked \cite{ParkerPR61}, thus coupled resonator systems based on rutile resonators are expected to be less sensitive as bolometers.

When the STO puck is radially offset, mode splitting occurs at $\varepsilon_r=230$, and the interaction between the two modes intensifies as the radial offset increases from zero. Level repulsion, defined as the distance between the two dispersion curves, increases with larger radial offsets, indicating stronger coupling between the two modes (Fig.~\ref{fig:CouplingEigenmode}(a)). This coupling causes the $Q$ factor of the STO-dominant mode (low $Q$ mode) to increase, while the $Q$ factor of the elliptical cavity-dominant mode (high $Q$ mode) decreases, and both converge at $\varepsilon_r=230$ (Fig.~\ref{fig:CouplingEigenmode}(b)). For a PEC elliptical cavity and an STO puck with $r_\mathrm{offset}$=\unit[50]{mm} and $\varepsilon_r=230$, the two modes exhibit equal quality factors of approximately $Q=2\times10^4$. One mode has a frequency of \unit[1279]{MHz}, while the other is at \unit[1315]{MHz}. The field plots for these two modes are presented in Fig.~\ref{fig:CouplingEigenmode}(c). In addition to changes in $\varepsilon_r$, the frequency sensitivities of the eigenmodes and their quality factors also depend on the radial offset. As shown in Fig.~\ref{fig:OffsetSensitivity}, at $\varepsilon_r$ = 230, where the coupling is at its maximum, even a slight radial offset of the puck results in a sharp reduction in the quality factor of the elliptical cavity-dominant mode. This is followed by an increase in the frequency sensitivity of the mode to changes in $\varepsilon_r$.

In practice, eigenmode information is obtained by measuring the S-parameters. Figure~\ref{fig:CouplingSParameters} illustrates the S-parameters for an excitation scheme using two antennas (as shown in Fig.~\ref{fig:SimulationSetup}), with a puck offset of \unit[50]{mm}. Two key parameters in the measurement are the frequency sensitivity of the modes to variations in $\varepsilon_r$ and the $Q$ factor of the modes. The quality factor can be calculated using a \unit[3]{dB} method and is also related to phase variations in the $S_{1,2}$ curve; larger phase changes (i.e., a steeper phase derivative) correspond to a higher $Q$ factor. Using the phase derivative offers the benefit of not only analyzing the peaks but also assessing the sensitivity of the transmission zero. This notch, located between two resonance peaks in the S-parameters, corresponds to a point of minimal signal transmission.

Figure~\ref{fig:CouplingSParameters}(c) illustrates the phase derivatives of the PEC cavity at various values of $\varepsilon_\mathrm{r}$. The distribution or pattern of the phase derivative near the resonance frequencies exhibits a Lorentzian shape. For modes near \unit[1.3]{GHz} (corresponding to the elliptical cavity's dominant mode), the phase derivative amplitude shows high values, indicating a large $Q$ factor. For the notch mode, the phase derivative peak remains nearly constant across different values of $\varepsilon_r$. For a normal conducting cavity, such as copper, the $S$-parameters are similar; however, the phase derivative amplitudes at the peaks are smaller, indicating a lower $Q$ factor.

Figure~\ref{fig:CouplingAntennaDerivatives} presents the frequencies of the two peaks in the $S_{1,2}$ curve, along with the notch between them. The resonance behavior as a function of $\varepsilon_r$ resembles eigenmode analysis, with level repulsion influenced by $r_\mathrm{offset}$. The frequency of the notch shifts similarly to that of the single STO puck calculated from the semi-analytical Eq.~\ref{AnalyticFormula}. The frequency sensitivity of modes 1 and 2 is generally lower than that of the notch, as shown in Fig.~\ref{fig:CouplingAntennaDerivatives}(b). 
Similarly to the eigenmode analysis, for $\varepsilon_r$ below 230, mode 1 is the elliptical cavity-dominant mode, whereas above 230, it becomes the second mode. For the elliptical cavity-dominant mode, frequency sensitivity peaks at $\varepsilon_r=230$, although this is accompanied by a decrease in its phase derivative amplitude that is related to $Q$ factor value. However, the product of frequency sensitivity and the phase derivative for the two modes is equal to that of the notch for a PEC cavity (Fig.~\ref{fig:CouplingAntennaDerivatives}(c)), whereas for a copper cavity (Fig.~\ref{fig:CouplingAntennaDerivatives}(d)), this product for the notch exceeds that of the two modes. This coupling allows for a trade-off between higher frequency sensitivity and a higher $Q$ factor, depending on the $\varepsilon_r$ value.

    		\begin{figure}[htp]
		\centering
        \subfloat[]{\includegraphics[width=1\columnwidth]{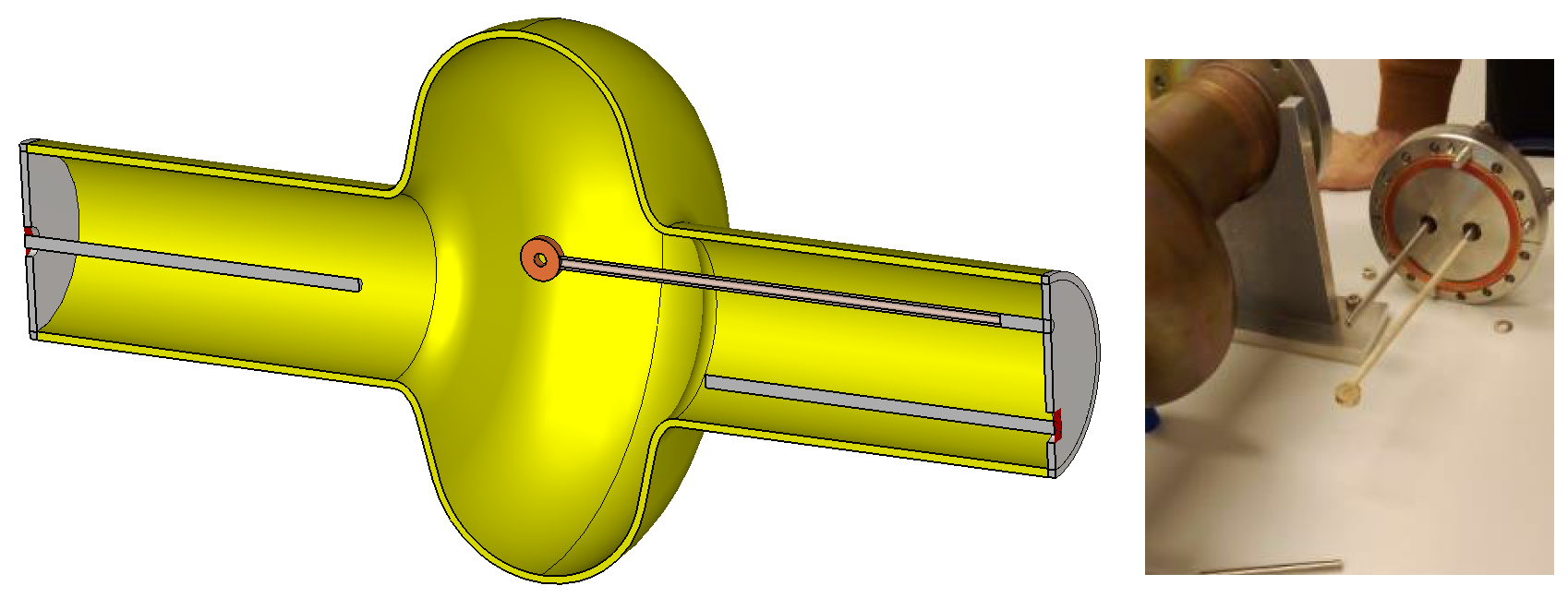}}\hfill
        \subfloat[]{\includegraphics[width=1\columnwidth]{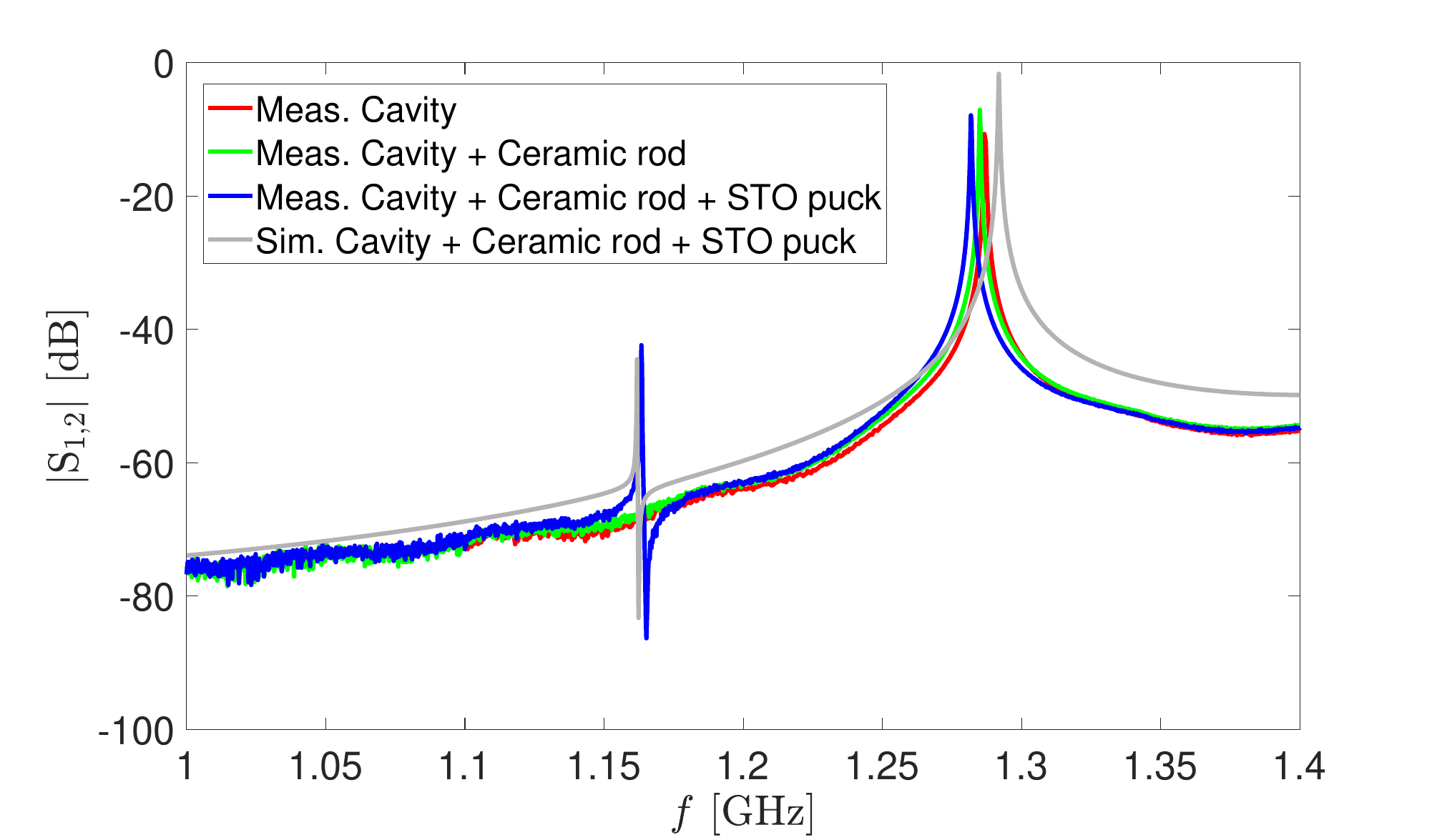}}\\
       \subfloat[]
        {\includegraphics[width=1\columnwidth]{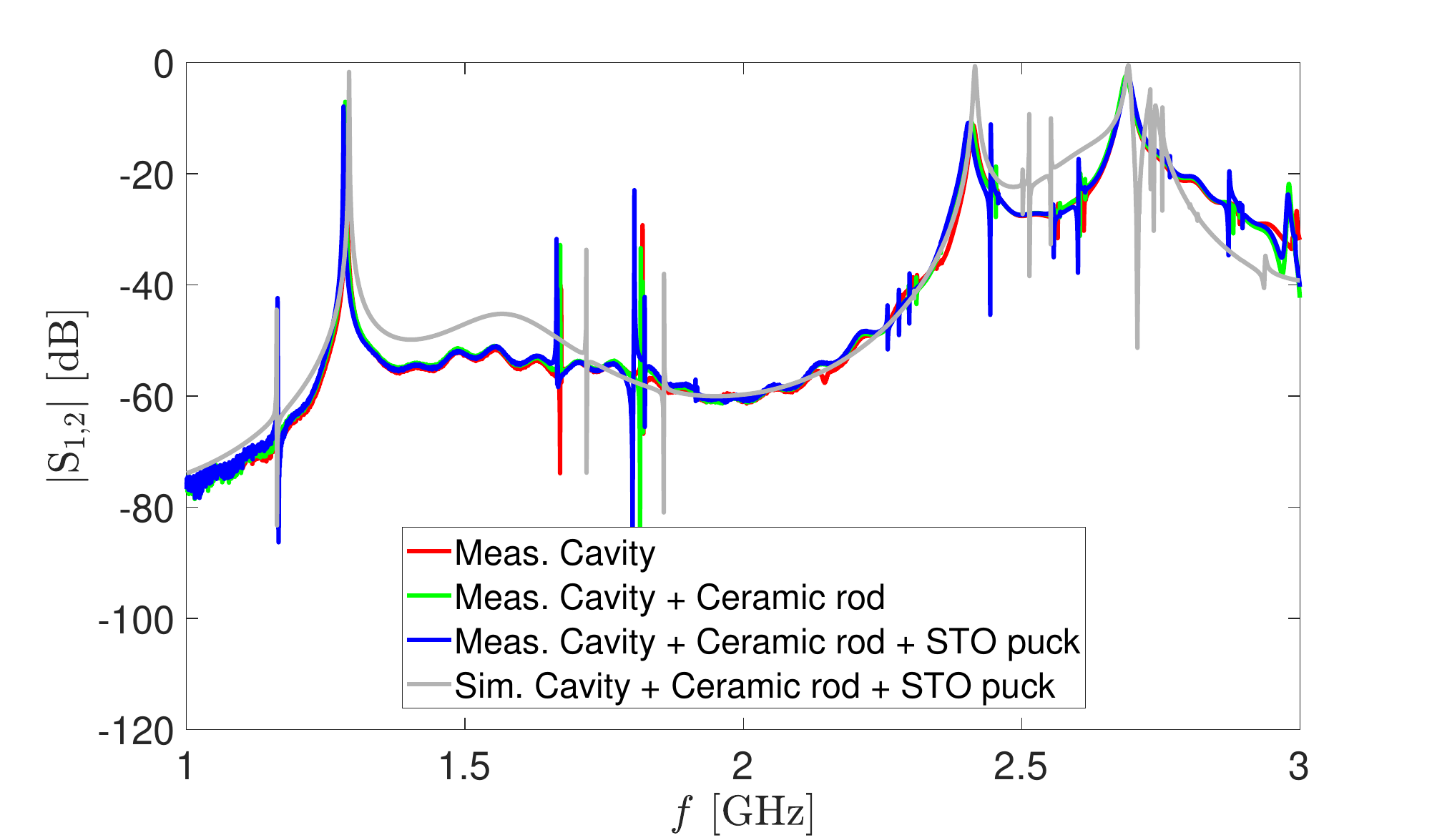}}\\
		\caption{The measurement setup (a) and measurement results of the coupled elliptical cavity with the STO puck over a narrow frequency range (b) and a wide frequency range (c). The STO puck is fixed to the flange via a long ceramic rod at $r_\mathrm{offset}$=\unit[19]{mm}. Two long antennas were used for the $S_{2,1}$ measurement: one at the flange center and the other at $r_\mathrm{offset}$=\unit[-19]{mm}. Note that the elliptical cavity used for the measurements had a slightly different shape than the one used for the simulations, causing a small change of the resonance modes of the cavity. An $\varepsilon_\mathrm{r}$ of 300 is assumed in the simulation.}
		\label{fig:MeasurementCavity}
	\end{figure}

\paragraph*{Test measurements.} To assess the coupling between the two modes in practice, a test case was established by placing the STO puck inside an elliptical cavity. Excitation is provided through two long rods placed at the end flanges, and the STO puck is secured at the center of the cavity with a radial offset of \unit[19]{mm}, mounted on a ceramic rod (see Fig.~\ref{fig:MeasurementCavity}). The resonance of the STO puck, observed at \unit[1.16]{GHz} in Fig.~\ref{fig:MeasurementCavity}(b), is consistent with the predictions of the simulations, corroborating the analysis reported in this section. 
In addition to the fundamental mode of the STO puck and the elliptical cavity, many other higher-order modes can also be excited in the resonators, which are partially reproduced by the simulated model (Fig.~\ref{fig:MeasurementCavity}(c)). Depending on the application, one or more of these modes with different frequencies, field patterns, and frequency sensitivities could also be selected for mode coupling.

    \section{Discussion}
	\label{sec:Discussion}

Unlike the power detection scheme, the fundamental limit in frequency measurements comes from the best achievable frequency stability of the measurement parts, and thus it is not directly related to Nyquist noise and ambient temperature, although cryogenic cooling may significantly improve the detection limit by increasing the quality factors. Using superconducting cavities and sapphire, it is possible to achieve $Q$ factors in the $10^9$ range \cite{Mann} and bandwidths of the order of 1~Hz at microwave frequencies \cite{Pischalnikov19, RomanenkoPRL23}. This, also considering that the resolution of the resonator linewidth is typically limited, by electronic noise, to about 1 ppm~\cite{Mann}, explains the unprecedented sensitivity of the frequency measurement approach.

In the coupled system, the high-$Q$ cavity can be used as a stable frequency reference enabling high-precision measurements, while the coupling with the STO resonator adds tunability and sensitivity to external physical quantities (Sect.~\ref{sec:CoupledCavity}). Although the specific characteristics of a particular sensor depend on its geometry, as well as by temperature and working conditions, we can envisage the potential of the coupled system in sensing applications. The maximum sensitivity of the frequency of the coupled system with respect to temperature occurs at resonance and reads 
\begin{equation}
    \left.\frac{\mathrm{d}f(T)}{\mathrm{d}T}\right|_\mathrm{max}=\beta \frac{\mathrm{d}f_\mathrm{STO}}{\mathrm{d}T}, 
    \label{temp_sensitivity}
\end{equation}
being $\beta$ a coefficient related to the coupling strength and the quality factor of the modes \cite{HaoIEEE03, GallopIEEE01}. The sensitivity is thus related to the temperature dependence of the STO frequency ($\mathrm{d}f_\mathrm{STO}/\mathrm{d}T$), and weighted by the coefficient $\beta$. 

In the bolometer detector described in \cite{GallopIEEE01, HaoIEEE03, Hao05, Hao20, Hao21}, two dielectric resonators, made using Sapphire and STO pucks, are coupled by fringing fields. In our experiment, the Sapphire resonator is replaced with the elliptical cavity that allows a better control of the coupling strength by varying the position of the STO puck within the cavity (Fig.~\ref{fig:CouplingEigenmode}). A simple expression of the maximum sensitivity of the coupled system (Eq.~\ref{temp_sensitivity}) can be obtained using the coupled-mode theory \cite{ElnaggarJMR14_1, ElnaggarJMR14_2} under resonant conditions. Considering the frequencies of mode 1 ($f_1$) and mode 2 ($f_2$) and their derivative with respect to temperature, $\mathrm{d}f_\mathrm{1(2)}/\mathrm{d}T=\beta_\mathrm{1(2)}(\mathrm{d}f_\mathrm{STO}/\mathrm{d}T)$, the $\beta_\mathrm{1}$ and $\beta_\mathrm{2}$ coefficients read (Appendix~\ref{App:CMT})
\begin{equation}
\beta_\mathrm{1}=\frac{\sqrt{1 - |\kappa|}}{1+\frac{Q_\mathrm{STO}}{Q_\mathrm{cav}}}
\label{eq:beta1}
\end{equation}
and
\begin{equation}
\beta_\mathrm{2}=\frac{\sqrt{1 + |\kappa|}}{1+\frac{Q_\mathrm{STO}}{Q_\mathrm{cav}}},
\label{eq:beta2}
\end{equation}
where the quality factors $Q_\mathrm{STO}$ and $Q_\mathrm{cav}$ refer to, respectively, the unperturbed dielectric resonator and cavity, and the coupling coefficient $\kappa$ is the normalized overlap integral of the electric fields of the dielectric resonator and cavity \cite{ElnaggarJMR14_1}. We note that the temperature sensitivity is improved using mode 2 at high $\kappa$, provided that $Q_\mathrm{cav} \gg Q_\mathrm{STO}$ (Eq.~\ref{eq:beta1}, \ref{eq:beta2}). From the simulated curves in Fig.~\ref{fig:CouplingEigenmode}(a) we derive $\kappa$ considering that, on resonance ($\epsilon_r=230$), $f_{1}=f_\mathrm{cav}\sqrt{1-\kappa}$ and $f_{2}=f_\mathrm{cav}\sqrt{1+\kappa}$ \cite{ElnaggarJMR14_2}. At $r_\mathrm{offset}$=\unit[10]{mm} and $r_\mathrm{offset}$=\unit[50]{mm}, the coupling coefficients are respectively $\kappa = 0.006$ and $\kappa = 0.03$, which give rise to $\beta_2 \approx 1$. Higher values of $\beta_2$ can be obtained by further improving the coupling strength. Being $\kappa$ a normalized quantity, Eq.~\ref{eq:beta2} suggests that the upper limit of $\beta_2$ is $\sqrt{2}$.

The frequency of the STO resonator is dependent on the permittivity $\varepsilon_\mathrm{r}$, which has a remarkable temperature dependence. From Eq.~\ref{AnalyticFormula} we obtain  
\begin{equation}
    \frac{\mathrm{d}f_\mathrm{STO}}{\mathrm{d}T}=-\frac{\alpha}{2} \varepsilon_\mathrm{r}^{-3/2} \frac{\mathrm{d}\varepsilon_\mathrm{r}}{\mathrm{d}T},
    \label{freq_derivative}
\end{equation}
being $\alpha$ related to the dimensions of the STO puck. As shown by electromagnetic simulations (Sect.~\ref{sec:CoupledCavity}), $\mathrm{d}f/\mathrm{d}\varepsilon_\mathrm{r}=2.8~\unit{MHz}$ for $\varepsilon_\mathrm{r} \approx 230$. At lower temperature, STO displays the structural transition around 100~K, the ferroelectric phase transition at 51~K and the stabilization of the quantum paraelectric phase below 5~K  \cite{GeyerJAP05, ZhaoIEEE22, HosainJAP19}. Between 20 and 80~K the relative permittivity of STO can be approximated as $\varepsilon_\mathrm{r} \approx 10^5/T \approx 2000$~\cite{GallopIEEE01}. Although in this range the $Q$ factor of the normal conductor cavity is typically lower than that of the sapphire resonator \cite{diVoraPRAppl22}, simulations suggest that, to obtain the maximum sensitivity, the change in the notch frequency should be considered (Fig.~\ref{fig:CouplingAntennaDerivatives}(d)). Below 9.2~K a Nb-coated TESLA cavity is superconducting and the $Q$ factor of the fundamental mode increases above $10^9$ \cite{TESLA00}. In optimal conditions, the superconducting cavity is cooled below the critical temperature while the STO puck is heated to $T \approx \unit[25]{K}$ to display the maximum $\mathrm{d}f_\mathrm{STO}/\mathrm{d}T$ \cite{Hao21}. However, at temperatures between 1 and 0.4~K the frequency of the STO resonator decreases monotonically with temperature (Fig.~\ref{fig:STO_low_temp}). From the fit, we derive $\mathrm{d}f_\mathrm{STO}/\mathrm{d}T=\unit[-10]{kHz/K}$ at \unit[0.8]{K}. This number is much smaller than what was obtained at $T\approx \unit[20-25]{K}$ \cite{Hao21}, yet it is expected to increase if small STO pucks are used (Eq.~\ref{freq_derivative}), as required, when the permittivity increases to $\varepsilon_r \approx 30000$, to match the frequencies of the STO resonator and cavity. Given $\varepsilon_\mathrm{r} \approx 30000$, we expect that resonators having a typical diameter in the range of  1~mm  are needed to achieve a significant resonant coupling with the 1.3 GHz cavity. Considering also that the heat capacity of the puck, which is directly related to the mass, plays a significant role in the effectiveness of the system as a bolometer \cite{HaoIEEE03}, we expect that reducing the size of the puck increases sensitivity by amplifying the temperature increase per unit of absorbed energy, thus enhancing the detection capabilities of the system.

Given the very peculiar characteristic of STO and the dependence of the STO resonator frequency by external factors such as applied electric field or mechanical stress, we just mention that possible applications of the proposed system of coupled cavity and STO puck are not limited to the realization of bolometers. Simulations indicate that the two cavities can be tuned to reach the strong coupling regime (Fig.~\ref{fig:CouplingEigenmode}). As expected, at resonance, the quality factor of the superconducting cavity progressively decreases as the coupling increases. However, we note that eigenmodes that have a $Q$ factor in the $10^5-10^6$ range can still be obtained by adjusting $r_\mathrm{offset}$ and the detuning of the natural frequencies of the superconducting cavity and the STO resonator.
These values are corroborated by transmission spectra measured at low temperature (Fig.~\ref{fig:STO_low_temp}), which show that the $Q$ factor of the STO resonator is in the $10^4$ range. This value can potentially be further improved by using opportunely prepared STO pucks. Electromagnetic simulations also show that the change in the frequency of the superconducting cavity mode and of the corresponding phase is amplified as an effect of the coupling with the STO resonator (Fig.~\ref{fig:CouplingSParameters}). In this regard, the phase derivative of $S_{1,2}$ shows a particularly steep response. 

We also mention that preliminary tests were carried out to test the stability of the STO mode frequency. In the case of measurements running for hours, the observed drift of the resonance frequency, due to external factors and without any feedback control, was in the range of 50~Hz both for the STO puck measured at 0.1~K and for superconducting cavities measured at 1.7~K. These values are strongly reduced, of about one order of magnitude, for timescales of a few seconds or minutes. By improving the experimental setup, and introducing a reference phase-locked loop (PLL) frequency stabilization, a significant reduction of the frequency drift is expected, thus allowing one to exploit phase measurements and further improve the system sensitivity. Additionally, the frequency tuning of the detection cavity can be ensured by the ferroelectricity of the STO and can be obtained by applying a DC electric field to the dielectric resonator. The expected tuning range is on the order of a few percent. 

	\section{Conclusions}
	\label{sec:Conclusion}

To conclude, we have studied the hybrid system composed by a high-$Q$ TESLA-shaped elliptical cavity and STO resonator, and investigated the effect of parameters, such as STO permittivity, puck dimensions and position within the cavity, that govern the coupling between the electromagnetic modes. Finite element simulations show that the hybrid system offers great versatility in tuning the coupling strength and achieving the strong coupling regime. These results are supported by test measurements carried out at room temperature using a copper cavity and a STO puck, and by the low-temperature characterization of the STO resonator, which shows a resonant frequency of 0.1 GHz and $Q$ factor of 10000 at 160~mK. The hybrid system shows potential for the realization of microwave sensors in which the sensitivity of the STO puck to selected physical quantities is exploited as the active element, while frequency or phase measurements in the high-$Q$ cavity are used to efficiently detect such changes. Our results are useful in designing dedicated experiments at low temperature, allowing the direct test of sensitivity and tunability of the proposed hybrid system in high-precision frequency-domain measurements.


\begin{acknowledgments}
We thank Fritz Casper, Sergio Calatroni, Akira Miyazaki, Walter Venturini, Alick Mcpherson, Giovanni Carugno and Marco Affronte for stimulating discussions. A.C. acknowledges financial support from PNRR MUR project E63C22002190007, PE23, NQSTI. A.G. acknowledges financial support from PNRR MUR project ECS\_00000033\_ECOSISTER. S.P. acknowledges financial support from PNRR MUR NQSTI PE000000023 project HIQRES CUP H43C22000870001.
\end{acknowledgments}

\appendix
\section{Temperature sensitivity of the coupled system}
\label{App:CMT}

We apply the coupled-mode theory (CMT) \cite{ElnaggarJMR14_1, ElnaggarJMR14_2} to derive a simple expression for the sensitivity of the coupled cavity-STO resonator system as a function of temperature. The components of the electromagnetic field are expressed as a linear superposition of the uncoupled ones, i.e.~the electric field of the coupled system is $\mathbf{E}=a_1\mathbf{E}_\mathrm{STO}+a_2\mathbf{E}_\mathrm{cav}$, being $\mathbf{E}_\mathrm{STO}$ the electric field of the dielectric resonator,  $\mathbf{E}_\mathrm{cav}$ the electric field of the cavity and $a_1, a_2$ the coupling coefficients. Similarly, the magnetic field is $\mathbf{H}=b_1\mathbf{H}_\mathrm{STO}+b_2\mathbf{H}_\mathrm{cav}$.

We assume that the fields are normalized ($\int_V \epsilon \mathbf{E}_\mathrm{STO} \cdot  \mathbf{E}_\mathrm{STO} dV=\int_V \epsilon \mathbf{E}_\mathrm{cav} \cdot  \mathbf{E}_\mathrm{cav} dV=A$), and consider the frequency ($f_{1}$) and quality factor ($Q_{1}$) of mode 1, and the frequency ($f_{2}$) and quality factor ($Q_{2}$) of mode 2. The ratio between coupled mode frequency and quality factor is \cite{ElnaggarJMR14_2}
\begin{equation}
\frac{f_{1(2)}}{Q_{1(2)}}=\frac{|a_1|^2f_\mathrm{STO}}{Q_\mathrm{STO}}+\frac{|b_\mathrm{2}|^2f_\mathrm{cav}}{Q_\mathrm{cav}}
\end{equation}
where $f_\mathrm{STO}$ and $Q_\mathrm{STO}$ are respectively the frequency and quality factor of the uncoupled dielectric resonator, while $f_\mathrm{cav}$ and $Q_\mathrm{cav}$ are the frequency and quality factor of the uncoupled cavity, respectively.

We also assume that the frequency of the dielectric resonator depends on temperature ($f_\mathrm{STO}=f_\mathrm{STO}(T)$), while the quality factors are, to the first approximation, temperature independent at least in a small temperature interval around the working point. The derivative with respect to temperature is
\begin{equation}
\frac{\mathrm{d} f_{1(2)}}{\mathrm{d}T}=|a_1|^2\frac{Q_{1(2)}}{Q_\mathrm{STO}}\frac{\mathrm{d} f_\mathrm{STO}(T)}{\mathrm{d}T};
\end{equation}
on resonance ($f_\mathrm{STO}=f_\mathrm{cav}$), the coefficients are $|a_1|=|a_2|=|b_1|=|b_2|=1/\sqrt{2}$ \cite{ElnaggarJMR14_2}. 

We consider the nonzero coupling coefficient $\kappa=\frac{\epsilon_0(\epsilon_r-1)}{A}\int_{V_\mathrm{STO}} \mathbf{E}_\mathrm{STO}^* \cdot  \mathbf{E}_\mathrm{cav} dV$, where $V_\mathrm{STO}$ is the volume of the dielectric resonator \cite{ElnaggarJMR14_1}. On resonance, the frequency and quality factor of the coupled modes are \cite{ElnaggarJMR14_2}: $f_{1}=f_\mathrm{cav}\sqrt{1-\kappa}$ and 
\begin{equation}
Q_{1}=2\sqrt{1-|\kappa|}\frac{Q_\mathrm{STO} Q_\mathrm{cav}}{Q_\mathrm{STO}+Q_\mathrm{cav}}
\end{equation}
for mode 1; $f_{2}=f_\mathrm{cav}\sqrt{1+\kappa}$ and
\begin{equation}
Q_{2}=2\sqrt{1+|\kappa|}\frac{Q_\mathrm{STO} Q_\mathrm{cav}}{Q_\mathrm{STO}+Q_\mathrm{cav}}
\end{equation}
for mode 2. The derivative with respect to temperature thus results 
\begin{equation}
\frac{\mathrm{d}f_{1}}{\mathrm{d}T}=\frac{\sqrt{1-|\kappa|}}{1+\frac{Q_\mathrm{STO}}{Q_\mathrm{cav}}} \frac{\mathrm{d}f_\mathrm{STO}}{\mathrm{d}T}
\label{eq:der_T_symm}
\end{equation}
for mode 1, and
\begin{equation}
\frac{\mathrm{d}f_{2}}{\mathrm{d}T}=\frac{\sqrt{1+|\kappa|}}{1+\frac{Q_\mathrm{STO}}{Q_\mathrm{cav}}} \frac{\mathrm{d}f_\mathrm{STO}}{\mathrm{d}T}
\label{eq:der_T_antisymm}
\end{equation}
for mode 2.

\end{document}